%% file: paper.tex
\documentclass[aps,prd,showpacs,preprintnumbers]{revtex4}
\usepackage{amsmath,amssymb}
\usepackage{color}
\usepackage{graphicx}
\usepackage{subfigure}
\usepackage{physics}
\usepackage{multirow}
\usepackage{axodraw2}
\input{macros.tex}

\begin{document}

\title{Interaction potential between heavy $\qqb$ in color octet configuration
  in QGP}
\author{Dibyendu Bala and Saumen Datta}
\email{dibyendu.bala@tifr.res.in, saumen@theory.tifr.res.in}
\affiliation{Department of Theoretical Physics, Tata Institute of Fundamental
Research, Homi Bhabha Road, Mumbai 400005, India.}

\begin{abstract}
  We investigate the interaction between a heavy quark-antiquark pair
  in color octet configuration in gluon plasma. We calculate
  nonperturbatively an effective thermal potential for such a pair
  through the study of the correlation function of a hybrid state with
  $\qqb$ octet and an adjoint gluon source in the static limit. We
  discuss the extraction of an octet potential, and present results
  for the effective thermal potential between octet $\qqb$ pair in
  gluon plasma for moderately high temperatures $\lesssim 2
  T_c$. The implications of our result are discussed.
\end{abstract}
\preprint{TIFR/TH/20-31}
\maketitle

\section{INTRODUCTION}
\label{sec.intro}
Quarkonia, mesonic bound states of heavy quark and antiquark,
provide one of the most important signatures of the formation of
quark-gluon plasma (QGP) in relativistic heavy ion collision
experiments. It was suggested in the early days of collider studies of
QGP \cite{satz} that the screening of the color charge inside QGP will lead to
dissolution of $J/\psi$ states, which can be observed by modification
of the dilepton peak. Various theoretical approaches have been
formulated to study the behavior of $\qqb$ bound states in the plasma.
See \cite{rothreview} for a recent comprehensive review \cite{review}.
In particular, an effective
thermal potential approach to the problem has been formulated
in \cite{impot} in perturbation theory, in the hard thermal loop (HTL)
approximation. The behavior of such a potential has been examined
in the effective field theory language for different distance scales
\cite{pnrT}. While the formulation of Ref. \cite{impot} is for the
theoretical abstraction of an isolated heavy $\qqb$ placed in the
plasma, the effective potential introduced there remains an important
part of a description of heavy quark systems in plasma using the open
quantum system language \cite{aka13,open}.  It is possible to evaluate
this potential nonperturbatively, from numerical calculations of
thermal Wilson loop \cite{rhs}.  Calculations based on Bayesian analysis
\cite{rhs,bkr} or modelling of the low-frequency peak \cite{pw}
suffer from large systematics. A new method, based on splitting of the
thermal Wilson loop, has led to a well-controlled extraction of the
potential from Wilson loop data \cite{singlet}.

For the phenomenology of quarkonia in plasma, one needs to know the
interaction between the $Q$ and the $\bar{Q}$ not only in the singlet
channel but also in the octet configuration.
The interaction of the $\qqb$
with the medium will change its color configuration from singlet to
octet and vice versa, and therefore study of evolution of the
$\qqb$ pair in plasma involves both color configurations.
In particular, in the open quantum system
approach developed to study quarkonia in medium \cite{aka13}, the
singlet and octet potentials are essential ingredients
\cite{open}. Also the $\qqb$ pair may be in a color octet state
at production time, 
and the time taken for it to go to a color
singlet combination may be larger than the formation time of the
plasma. In particular, this is expected to be the case for quarkonia
at large $p_\perp$ \cite{vitev}.
However, nonperturbative information about the octet
potential is not available in the literature. Part of the problem is
the inherent difficulty in defining an octet state in a
gauge-invariant set-up.

For the singlet, the potential describes the time evolution of the
thermal correlator of the nonrelativistic vector current, 
\beq
\cgt \equiv \int d^3x \, \left\langle \chi^\dagger\left(t,\vec{x}
\right) \sigma_k \; 
\wil^\dagger \psi\left(t,\vec{x}+\vec{r}\right) \
\psi^\dagger\left(0,\vec{x}+\vec{r}\right) \sigma_k \, 
\wil \, \chi \left(0,\vec{x}\right) \right\rangle \; \cdot
\eeq{nrc}
Here $\wil$ is a suitable gauge connection such that the current is
gauge invariant, $\psi, \ \chi$ are nonrelativistic fields that
annihilate a quark and create an antiquark, respectively, and the
angular brackets denote thermal average. The $\sigma_k$ do not affect
the $\ord(m_Q^0)$ potential.  If one has a system where the sole
interaction term is a potential $V(\vec{r})$ between the quark and the
antiquark, then it is easy to show that, to leading order in $1/\mq$,
$\cgt$ satisfies
\beq
\left( i \, \partial_t \, - \, \frac{\nabla_{\vec{r}}^2}{\mq} \right) \cgt
\ = \  V(\vec{r}) \; \cgt . 
\eeq{nr}
We then define a potential \cite{impot} by equating the left hand side
of \eqn{nr} to $V(t, \vec{r}) \; \cgt$ (staying within leading order
of $1/\mq$), where the interaction effects are summarized in a
time-dependent $V(t, \vec{r})$.  An effective thermal potential,
$\vt$, can then be defined in the large $t$ limit, if the limit
exists: $\vt = \lim_{t \to \infty} V(t, \vec{r})$. In the static
limit, modulo renormalization factor, $\cgt$ reduces to a
Minkowski-time Wilson loop:
\beq
W_M(t, \vec{r}) \ = \ \frac{1}{3} \, {\rm Tr} \;  \mathbb{P}
\, e^{i \int_0^t dt_1 A_0(t_1,\vec{r})} \; \wil\left(t; \vec{r},
  \vec{0}\right) \; \mathbb{P} \, e^{i \int_t^0 dt_2 A_0(t_2, \vec{0})}
  \wil\left(0; \vec{0},\vec{r} \right)
\eeq{wm}
and \eqn{nr} reduces to
\beq
i \, \partial_t \; \log W_M(t, \vec{r}) \xrightarrow[t \to \infty]{}
\vt, \eeq{pot}
which defines the effective thermal potential for the singlet channel
\cite{impot}. 

\eqn{nr} describes the time evolution of a thermal correlation
function, and not of a $\qqb$ wave function. It reduces naturally to
\eqn{pot} in the static limit, which can lead to a nonperturbative
calculation of the potential \cite{rhs}. The effective thermal
potential defined by \eqn{nr} is complex in general \cite{impot}. 
It has been argued that the
potential \eqn{nr} can be re-interpreted in terms of evolution of a
$\qqb$ wave function \cite{ar,kaar}. In this language, the
evolution of the $\qqb$ pair is described by a stochastic Hamiltonian,
and $\vim$ is related to fluctuation of the stochastic noise.
The effective thermal potential remains an essential ingredient in
such open quantum system studies of quarkonia in plasma.

For the octet potential, one can proceed in a similar way, starting
with a point-split nonrelativistic current
\beq
J^a(\vrr; \vx,\vxo, t) \ =
\ \psi^\dagger\left(\vec{x}+\vec{r}; t \right) \,
\sigma_k \, 
\wil(\vec{x}+\vec{r}, \vec{x}_0; t) \, T^a \, \wil(\vec{x}_0,
\vec{x};t) \, \chi \left(\vec{x}; t\right) \; \cdot
\eeq{octj}
and looking at the time derivative of the correlator $\cgto$
{\em a la} \eqn{pot}. The current $J^a$  is gauge dependent
and the correlator $\cgto$ needs to be defined in a fixed gauge.
Unfortunately, standard gauge fixed definitions of $\cgto$ lead to a
system which may be very different from what was intended: e.g., in
the temporal gauge $\cgto$ actually describes, in the static limit, a
$\bar{Q} Q^a_{\rm adj} Q$ system \cite{philipsen}. In the literature
one usually employs the Coulomb gauge; nonperturbatively, $\cgto$ is
not defined in the Coulomb gauge, and a further fixing
of temporal gauge along $\vec{x}_0$ gets us back to a $\bar{Q}
Q^a_{\rm adj} Q$ system \cite{philipsen}.

When we talk about $\bar{Q} Q$ in color octet combination in
the context of quarkonia, we have in mind a system where $\bar{Q} Q$
is interacting with an adjoint gluonic source. To mimic this system,
we could start with a trial current like 
\beq
J_G(\vrr; \vx, \vxo, t) \; = \; \bar{\psi}(\vx+\vrr; t) \,
\wil (\vx+\vrr, \vxo; t) \, G^a(\vxo; t) T^a \, \wil (\vxo, \vx; t)
\, \chi(\vx; t) \; ,
\eeq{hybridj}
which is a color singlet combination of the color-octet
quark-antiquark system and an adjoint gluonic source at a time slice
$t$, and then look at the correlator $C_G(t,\vec{r}) =
\langle J_G^\dag(t) J^G(0) \rangle$.
With a judicious choice of $G$, it is possible to ensure that $J_G$
does not have overlap with a configuration where the quark-antiquark
system is in color-singlet state.  A color singlet state consisting of
$\qqo$ and adjoint gluon source is called a hybrid state. At zero
temperature, hybrid potentials have been studied in detail in the
literature \cite{octet0}. For us, the important information is that in
certain regimes, the hybrid current can be used to define an octet
potential \cite{pnrpot,balipineda}.

In this work, we study the thermal effect on the hybrid Wilson loop
and extract information about the thermal modification of the
interaction potential between static $\qqb$ in color octet
configuration in gluon plasma. To our knowledge, this is the first
study of the effective thermal color octet potential, though a related
quantity, the color octet free energy, has been studied before
\cite{free,tumfree}. Preliminary results of this study were presented in
\cite{lattice}. The crux of the problem is to extract $\vt$; this is
discussed in \scn{analysis}.
We discuss our method in detail in
\scn{analysis}. Details of the numerical calculation are given in
\scn{lat}.  Our results for the potential are given in \scn{vpot}, and
in \scn{summary} we summarize and discuss the results. Some technical
details are relegated to the appendix: \apx{ape} discusses role of
smearing in our study, some details relevant for \scn{analysis} can be
found in \apx{lo}, and various systematics of our extraction of the
potential can be found in \apx{system}.

\section{Observables and Analysis}
\label{sec.analysis}
For our study, we use the hybrid current operator \eqn{hybridj} with
the chromomagnetic field operator for $G \, = \, G^a T^a$: we use
the two choices $\bz$ and $\bxy \, = \, B_x \, + \, i B_y$. Here $z$ is
taken to be the separation between the quark and the antiquark. In the
static limit, one gets Wilson loop with insertion of the $G$ field:
\beq
W_M^G(\tau; \vec{r}, \vec{x}) \ = \ \frac{1}{3} \, {\rm Tr} \;  \mathbb{P}
\, e^{i \int_0^\tau d\tau_1 A_0(\tau_1,\vec{r})} \; \wil\left(\tau; \vec{r},
\vec{x}\right) \; G^\dag(\tau; \vec{x}) \; \wil\left(\tau;
\vec{x}, \vec{0}\right) \;
\mathbb{P} \, e^{i \int_\tau^0 d\tau_2 A_0(\tau_2, \vec{0})}
\wil\left(0; \vec{0},\vec{x} \right) G(0; \vec{x}) \;
\wil\left(0; \vec{x}, \vec{r}\right) \, \cdot
\eeq{wmg}
In order to define a potential, we will need to go to Minkowski time
and take long time derivative, similar to \eqn{pot}.

While the aim of the paper here is to calculate the effective thermal
potential nonperturbatively, in order to understand the method,
it helps to think in terms of perturbation theory. 
In leading order (LO) of perturbation theory, the effect of the insertion
$G$ isolates and one gets the octet potential from the long time
behavior of Wilson loop:
\beq
  \raisebox{-19pt}{
    \begin{axopicture}(60,60)
    \EBox(0,0)(50,50)
    \GCirc(25,0){4}{0.80}
    \GCirc(25,50){4}{0.80}
  \end{axopicture}} \ \equiv \
    \raisebox{-19pt}{
  \begin{axopicture}(20,60)
    \GCirc(10,50){4}{0.80}
    \GCirc(10,0){4}{0.80}
    \DashDoubleLine(10,0)(10,50){1.5}{2}
  \end{axopicture}} \ \left( 1 \ + \ 
  \raisebox{-19pt}{
    \begin{axopicture}(60,60)
    \EBox(0,0)(50,50)
    \DashLine(0,25)(50,25){3}
    \GCirc(25,50){1.5}{1.0}
    \GCirc(25,0){1.5}{1.0}
  \end{axopicture}} \, + \,
  \raisebox{-19pt}{
  \begin{axopicture}(60,60)
    \EBox(0,0)(50,50)
    \DashArc(0,25)(10,-90,90){3}
    \GCirc(25,50){1.5}{1.0}
    \GCirc(25,0){1.5}{1.0}
  \end{axopicture}} \, + \, 
  \raisebox{-19pt}{
  \begin{axopicture}(60,60)
    \EBox(0,0)(50,50)
    \DashArc(50,25)(10,90,-90){3}
    \GCirc(25,50){1.5}{1.0}
    \GCirc(25,0){1.5}{1.0}
  \end{axopicture}} \, + \ \cdot \cdot \cdot \right)
  \eeq{factor}
Here time direction is shown vertically, the grey circles indicate the
magnetic fields, and the empty dots and dashed lines on the Wilson
loop indicate color matrix insertion $T^a$ and $00$ component of the
gluon propagator, respectively.  The leading order potential comes
from a ladder sum of diagrams like those explicitly shown inside the
parentheses in \eqn{factor}, where the effect of the $G$ insertion is
merely a change in color factor due to the color matrix insertions in
the Wilson loops. Such factorization will not hold 
nonperturbatively, where $J^G$ for the $\bz$ operator will give
rise to potential for the L=0 state $\Sigma_u^-$ and the
$\bxy$ operator, that for the L=1 state $\Pi_u$. Here L refers to
angular momentum around the axis joining the quark and antiquark, $u$
denotes CP odd, and $-$ refers to parity for reflection about a plane
passing through this axis. The
potential for these operators have been studied \cite{octet0}, and its
connection to the octet potential has been explored in detail
\cite{balipineda}. In the deconfined phase, such hybrid states are
not expected to survive. However, we will sometimes refer to the 
potentials obtained with the two operator insertions as L=0 and L=1
potentials, respectively. 

At short distances, we expect the potential for a state like
\eqn{hybridj} to give the potential for the color octet $\qqb$ state,
modulo a constant term capturing the effect of the $G$ insertion
\cite{pnrpot, balipineda} :
\beq
V_G(r) \ \sim \ V_O(r) \; + \; \Lambda_G \; + \; \ord(r^2)
\eeq{hybridG}
Clearly, for the two operator choices here, $\Lambda_G$ is
identical. So we expect $V_G$ to be same for the two operators in the
short distance regime, modulo $\ord(r^2)$ effects. This behavior was
tested in detail in \cite{balipineda}: while the convergence of $V_G$
extracted from the two components of the magnetic field was verified,
it becomes difficult to isolate the color octet potential due to the
quick onset of the nonperturbative effects.

At finite temperatures, we can define a thermal Wilson loop similar to
\eqn{wmg}, except the $\tau$ extent is now finite: $0 < \tau <
\beta=1/T$. Just as in the case of the singlet \cite{impot}, one can
define in HTL perturbation theory an effective
thermal potential by continuing to Minkowski time and taking a long
time derivative {\em a la} \eqn{pot}. For completeness, we outline the
steps in \apx{lo}. In this order, the factorization of \eqn{factor}
holds and we get, in the Debye screening regime, a thermal potential
\begin{eqnarray}
  \voc &=& \vro \; - \; i \, \vio \label{octlo} \\
  \vro &=& \frac{g^2}{2 N_c} \; \frac{e^{-\md r}}{4 \pi r} \ - \
  \frac{g^2 \, C_F}{4 \pi} \, \md \nonumber \\
\vio &=& \frac{g^2 T}{2 \pi} \left[ \frac{N_c}{2} \int_0^\infty
  \frac{dz \, z}{(z^2+1)^2} \ - \ \frac{1}{2 N_c} \int_0^\infty
  \frac{dz \, z}{(z^2+1)^2} \; \left( 1 \, - \, \frac{\sin zx}{zx} \right)
  \right] \nonumber
\end{eqnarray}
where $\md$, the Debye mass, $= g T$  in this order of
perturbation theory (for gluon plasma), and
$x = \md r$. 
It is interesting to compare it to the thermal singlet potential
\cite{impot}:
\begin{eqnarray}
\vre &=& - \frac{g^2 C_F}{4 \pi r} \; e^{-\md r} \ - \
\frac{g^2 \, C_F}{4 \pi} \, \md \nonumber \\
\vim &=& \frac{g^2 \, C_F \, T}{2 \pi} \int_0^\infty \frac{dz \, z}{(z^2+1)^2}
\, \left( 1 \, - \, \frac{\sin zx}{zx} \right) \label{impots} \ \cdot
\end{eqnarray}
Both $\vio$ and $\vim$  $\to T \, C_F \, g^2/4 \pi$ as $r \to \infty$.
$\vre$ corresponds to the usual
physics of Debye screening in medium, such that for sufficiently large
screening, the bound states will not form. On the other hand, $\vim$
clearly leads to a broadening of the spectral function peak. It
captures the physics of collision with the thermal particles leading
to a decoherence of the $\qqb$ wave function \cite{kaar,aka13}.

The expressions \eqn{octlo} and \eqn{impots} are valid only
in the HTL limit $T \gg 1/r$. For the distance regime $rT \ll 1$, the
thermal correction to the singlet potential has been calculated using
effective field theory techniques \cite{pnrT}. At short distances,
$\vim \sim r^2$ \cite{pnrT}. This is indeed the parametric behavior
seen in the short distance regime nonperturbatively \cite{singlet}.
\eqn{octlo} is not expected to give us the correct potential, even
qualitatively, at all distance scales; they, however, are compact and
are useful in understanding certain features of the thermal potential. 

Our aim here is to extract the potential nonperturbatively from the
Euclidean Wilson loop \eqn{wmg}. For the singlet, nonperturbative
assessment of $\vt$ has been done, e.g., in \cite{bkr,pw,prw}, and
recently in \cite{singlet}. The potential, in particular $\vim$, is
very different from \eqn{impots} in the temperature range $\lesssim
2T_c$. Here for the analysis of $\wmg$ we will follow the strategy of
\cite{singlet}, which we outline below. See \cite{singlet} for a more
detailed discussion.

At zero temperature, modulo renormalization factors, the Minkowski
space Wilson loop has the asymptotic time behavior $\wm \sim e^{-i
  V(r) t}$ (\eqn{pot}), leading to the Euclidean time behavior $\we
\sim e^{-V(r) \tau}$. Going to sufficiently long $\tau$, this behavior
is indicated by a plateau in $- \partial_\tau \log \we$, from which we
extract $V(r)$. At finite temperature, one does not see such a plateau
behavior. It was pointed out in \cite{singlet}, however, that
splitting the Wilson loop in parts ``symmetric'' and ``asymmetric''
around $\tau = \beta/2$,
\beq
\wap \ = \ \sqrt{\frac{\wrt}{\wrb}}, \qquad \qquad
\wpr \ = \ \sqrt{\wrt \times \wrb},
\eeq{wsplit}
one can extract a plateau structure from $\log \wap \approx (\beta/2 \,
- \, \tau) \; V_r$ (see also \scn{lat}). This behavior is illustrated
in the left panel of \fgn{combo}. 

\bef
\centerline{\includegraphics[width=7cm,height=6cm]{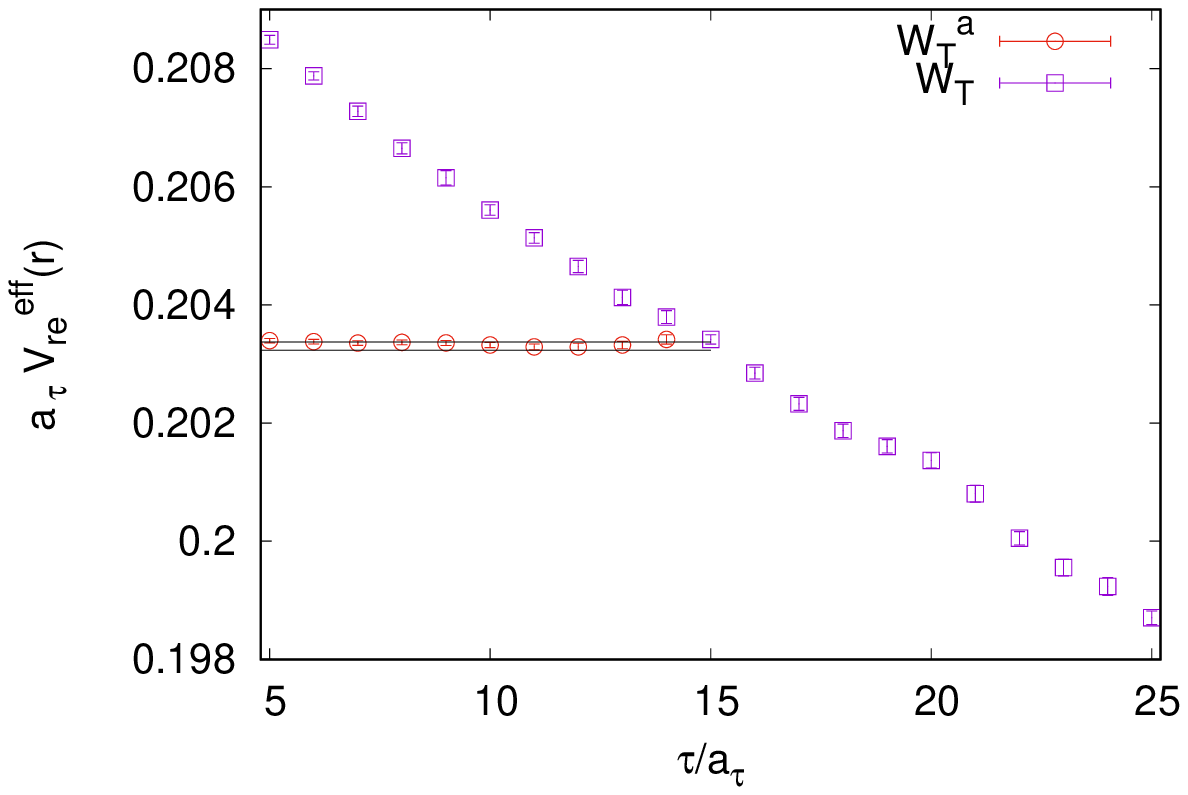}
\hspace{1cm}\includegraphics[width=6.5cm,height=5.5cm]{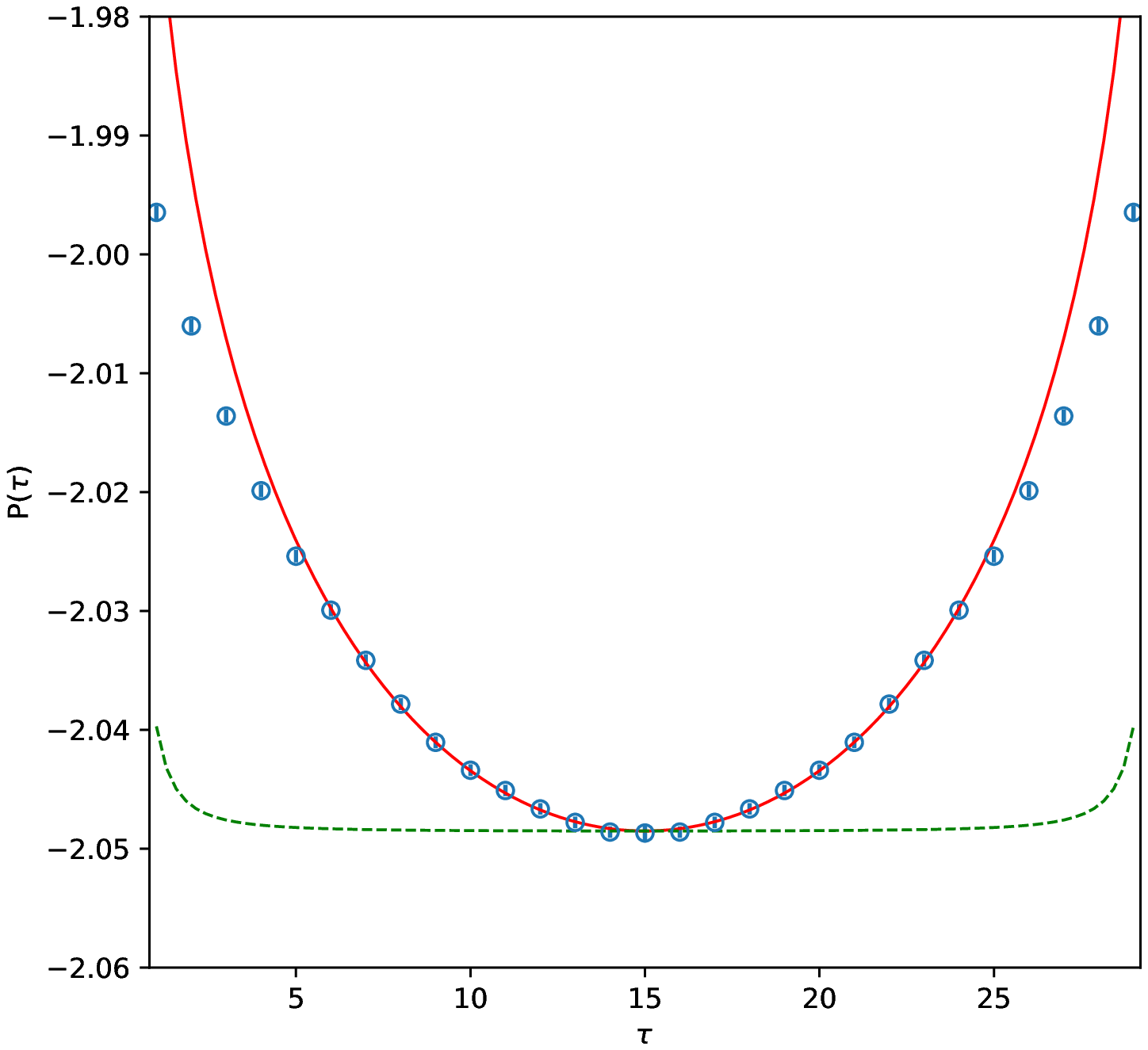}}
\caption{(Left) ``Local mass'' plot from $W^{a}_{T}$ and $W_{T}$ for
  Set 3, at 1.5 $T_c$, at $r/a_s=6$ (smearing level=200) for the
  Singlet. (Right) $P(\tau) = \log \wpr$, shown together with the
  value of the periodic part from \eqn{combo} (red, full) and the first term of
  \eqn{higher} (green, dashed).} 
\eef{combo}

The plateau from $\wap$ gives the real part of the potential, while
$\wpr$ contributes to the imaginary part. This can be understood by
writing a spectral decomposition for $\ptau \ = \log \wpr$:
\begin{eqnarray}
\ptau &=& \int_{- \infty}^\infty d\om \ \sw \ \frac{1}{2} \left(
e^{-\om \tau} \, + \, e^{-\om (\beta - \tau)} \right) \ \ + \,
{\rm \tau-independent \ terms} \nonumber \\
\tau \to i t \Rightarrow \ \ i \partial_t P(i t) &=&
\int_{- \infty}^\infty d\om \ \sw \ \frac{\om}{2} \left( e^{-i \om t} \, - \,
e^{-\om \beta} e^{i \om t} \right) . \label{argsplit} 
\end{eqnarray}
At large $t$, 
the oscillating factors $\exp(\pm i \om t)$ ensure that only the $\om \to 0$
contribution to the integral in \eqn{argsplit} survives. Then
$\exp(\beta \om) \to 1$ and \eqn{argsplit} shows that $P(i t)$ leads
to an imaginary potential.

To proceed further, we note that $\exp(-i
\, \om \, t) \, - \, \exp(i \, \om \, t \, - \, \om \, \beta)
\xrightarrow{t \to \infty} - 2 \pi \, i \, \omega \, \delta(\omega)$.
Then in order to get a finite potential
$-i \, \vim = \lim_{t \to \infty} i \, \partial_t \, P(it)$ we need 
$\sw \ \underset{\omega \to 0}{\thicksim} \ \frac{1}{\omega^2} \;
\left( 1 + \mathcal{O}(\omega) \right)$ . Since we expect thermal
physics to introduce a distribution function $(1 + \nbw)
\xrightarrow{\omega \to 0} \frac{\textstyle T}{\textstyle
\omega}$ in $P(\tau)$ (see \apx{lo}), the existence of
a potential requires a low $\om$
structure of $\sw \sim (1+\nbw) \ \frac{\textstyle \beta \,
  V_{\rm im}}{\textstyle \pi \, \omega} $. This 
leads to the following behavior of the Wilson loop near $\beta/2$:
\beq
\wrt \ = \ e^{- \vre \left(\tau \, - \, \frac{\beta}{2}\right) \; - \;
  \frac{\beta}{\pi} \, \vim \, \log  \sin \left(\frac{\pi \, \tau}{\beta}
  \right) - ....} \ W_{\scriptscriptstyle T}(\beta/2, \vec{r})
\eeq{combo}
where the higher order terms $....$ are non-potential terms. For the
periodic part, expanding $\sw \; \left( (1-\exp(-\beta \omega)\right)$
in series of $\omega$ will give \cite{singlet}
\beq
... \ = \ \sum_l c_l \; \frac{(2l \, - \, 1)!}{\beta^{2 l}} \; \left(
\zeta \left( 2 l, \, \frac{\tau}{\beta} \right) \; + \; \zeta \left( 2
l, \, 1 \, - \, \frac{\tau}{\beta} \right) \; - \; 2 \, \zeta
\left( 2 l, \, 0.5 \right) \right)
\eeq{higher}

Just the simple form \eqn{combo}, without any corrections, gives a
very good description of the Wilson loop data around
$\beta/2$. We show one illustration of this in \fgn{combo}. Here we
use \eqn{combo} with the first term of \eqn{higher} to fit the singlet
data. In the left plot we show the fitted value for $\vre$ on top of
the 'local values' obtained from $\wap$. In the right panel we show
$P(\tau)$, defined above \eqn{argsplit}, along with the contribution
of the $\vim$ term in \eqn{combo} and that of the $c_1$ term. The
$\vim$ term captures the behavior of the data near $\beta/2$ very
well. The first term of \eqn{higher} has a very different behavior
near the center. Its addition does not significantly improve the fit
quality ($\chi^2$); however, it can sometimes destabilize the plot if the
interval is small, or if the data is not very accurate, as is often
the case for $\wmg$. Therefore for the octet, we stick to just the
form in \eqn{combo}, and choose a suitable interval so that the fit
quality is good. In \scn{vpot} we will show how well \eqn{combo}
explains the
data, by examining plateaus for the local values of the potential.

The discussion of the potential requires a low-$\omega$ peak. In order
to successfully determine the potential, it is necessary to get a
region in $\tau$ where this peak dominates the contribution to the
potential. In earlier literature, this peak has been hypothesized to have a
Lorentzian or Gaussian structure. However, the form in \eqn{combo} leads
to an asymmetric peak: while for $\omega \approx V_r$ it gives a
Breit-Wigner structure, the fall-off from the peak is very different
in the large-$\omega$ and small-$\omega$ side. The limiting behaviors are
\begin{align} 
\rlow &\approx \sqrt{\frac{2}{\pi}} \ \frac{\viss}{(\vrss - \omega)^2 \; +
    \; \viss^2} \qquad &{} |\vrss - \omega|, \; \viss \ll T \nonumber \\
  & \sim  (\omega - \vrss)^{- \left(1-\frac{\beta \viss}{\pi} \right)} &{} \omega -
  \vrss \gg T \label{asymp} \\ 
  & \sim  e^{- \beta (\vrss - \omega)} \ (\vrss -
  \omega)^{-\left(1-\frac{\beta \viss}{\pi} \right)} &{} \omega - \vrss \ll -T \nonumber
\end{align}
  
An asymmetric peak structure has also been suggested in Ref. \cite{br1},
where the spectral function for thin Wilson loops was discussed from HTL
perturbation theory. It agrees with \eqn{asymp} near the peak, but
starts disagreeing away from it. The peak in \eqn{asymp} takes into
account that the non-potential modes have been sufficiently suppressed
by smearing, so that only the potential part \eqn{combo}, contributes,
as is supported by the data. 

\section{Details of the Calculation}
\label{sec.lat}

As mentioned in the previous section, our primary quantity is the
nonperturbatively estimated value of the Wilson loop $\wmg$, 
\eqn{wmg}, in a gluonic plasma, at moderately high temperatures
$\lesssim 2 \tc$.  Since we need a very fine grid of points for the
extraction of the potential from the Wilson loop, we have used a
space-time anisotropic discretization, with $\xi \; = \; \as/\at \; =
\; 3$.  We have generated lattices with the anisotropic Wilson action
\beq
S_W \ = \ - 3 V \left(\bs \, P_s \; + \; \bt \, P_t \right),
\qquad {\rm where}
\ P_s \; = \; \frac{1}{3 V N_c} \sum_x \sum_{i < j} \, {\rm Re} \,
   {\rm Tr} \; U_{ij}(x), \quad \quad P_t \; = \; \frac{1}{3 V N_c}
   \sum_x \sum_i \, {\rm Re} \, {\rm Tr} \; U_{0i}(x)
\eeq{plaq}
are the spatial and temporal plaquette variables, $V=N_s^3 \, \times
\, \nt$, $x$ runs over all space-time points, $i$ runs over all
spatial indices, and $U_{mn}(x)$ is the gauge connection around the
$mn$ direction plaquette at $x$.  We follow the method of Ref.
\cite{klassen} to get $\bs$, $\bt$; $\xi$ is estimated from comparison
of spatial and temporal Wilson loops.  The lattice parameters we use
are given in \tbn{sets}.  For each set, short Monte Carlo runs are
made at closely spaced $\nt$ to find the $\nt$ for deconfinement
transition, and then other temperatures are obtained by varying $\nt$
. The value of $\nt$ at $T_c$, determined by the peak of the Polyakov
loop susceptibility, is shown in \tbn{sets}. For the measurements we
ran multiple Monte Carlo chains: first \#init number of
well-decorrelated, thermalized configurations were generated. From
each of these starting configurations long Monte Carlo chains were
constructed: each chain was further thermalized, and then \#meas
measurements made, two measurements being separated by (100 timelink
multilevel \cite{multilevel} hits + 100 update sweeps).  Each update
step consisted of 1 heatbath + 3 overrelaxation steps, whereas for the
multilevel only heatbath steps were used.

We therefore have \#init $\times$ \#meas measurements of timelike
Wilson loops, i.e., the Euclidean time version of $W_M$, with magnetic
field insertions (\eqn{hybridj}) at the middle of the spatial
connection. All averages and errors are done via a bootstrap analysis;
for the bootstrap, we have further averaged the $\ord(10^3)$
measurements into $\ord(10^2)$ blocks ($\ord(10)$ for set 1). The
magnetic field operators have been implemented using the clover
construction.  As in the singlet case \cite{singlet}, we do APE
smearing \cite{ape} of the spatial links to reduce the non-potential
effects.  At each APE step, a spatial gauge link is replaced by $ {\rm
  Proj}_{SU(3)} \, \left( \alpha \times \; {\rm link} \, + \, \sum
\ {\rm spatial} \ {\rm staples} \right)$, where we kept $\alpha$ =
2.5. We have looked at data from a number of APE steps (shown in
\tbn{sets}). With higher number of APE steps, the data quality
decreases, but the effect of the non-potential terms also
decreases, making it easier to reach a plateau and extract a potential. 

\bet
\caption{List of lattices used in our calculation. For each set,
  \#meas Wilson loop measurements were done for each of \#init
  parallel runs, giving a statistics of \#meas $\times$ \#init. See
  text for other details.}
\setlength{\tabcolsep}{8pt}
\begin{tabular}{lllccccl}
  \hline
  set & $\bs$ & $\bt$ & $\nt (\tc)$ & size & \#init & \#meas & Smearing \\
  \hline
  1 & 2.53 & 15.95 & 30 & $48^3\times 20$ & 47 & 25 & 200 \\
  \hline
  2 & 2.57 & 16.53 & 38 & $32^3\times 25$ & 91 & 100 & 150,
  200,250,300 \\
  & & & & $32^3 \times 32$ & 91 & 100 & 150, 200, 250, 300 \\
  \hline
  3 & 2.60 & 16.98 & 45 & $40^3\times 23$ & 91 & 90 & 150, 200, 250,
  300, 400 \\
  & & & & $40^3\times30$ & 91  &  90 & 150, 200, 250, 300, 400 \\
  & & & & $40^3\times38$ & 89 & 90 & 150, 200, 250, 300, 400 \\
  & & & & $30^3\times60$ & 91 & 140 & 200 \\
  \hline
\end{tabular}
\eet{sets}

The use of spatial gauge link smearing is well-known in potential
studies, and detailed nonperturbative studies exist. In our context,
it is instructive, however, to understand its effect in the leading
order for the Wilson loop. this is discussed in \apx{ape}. 
Some illustrations of its effect on the extracted potential can be
found in \apx{system}. 

In this work, we have given all physical
quantities in temperature units. Conversion to physical units, if
needed, can be done by setting $T_c$ to 280 MeV, which is the value
obtained by fixing the string tension: $\sqrt{\sigma}$ = 0.44 GeV.
The spatial size of the lattices are $> 1.8$ fm, with the lattice of
set 1 being 3.4 fm.

\section{$\wmg$ and the octet $\qqb$ interaction potential}
\label{sec.vpot}
Before presenting the results for the effective potential for
$C_G(t,\vec{r})$, \eqn{hybridj}, we illustrate how
well the form \eqn{combo} explains the data, by doing the equivalent of
a local mass plot: we extract the
``local potential'' from a subset of data points. We find $V(r;
\tau)$ from the data $W(r; t)$ with $t=\tau, \tau+1, N_t-\tau,
N_t-\tau-1$. If the data is dominated by the potential term near
$\beta/2$, we will expect a plateau near $\beta/2$. We give two
examples of such an effective potential plateau in \fgn{plateau}.
To get the results, we have done a bootstrap analysis, where the
parameter values within each bootstrap sample are obtained by a
$\chi^2$ fit with the full covariance matrix.
The data is seen to show a plateau behavior in a region around
$\beta/2$. The final value of the potential (within each bootstrap
sample) is then obtained by doing a
fit to \eqn{combo} over this plateau range. The statistical error
for a given fit range is the (16,84) percentile band of the bootstrap
distribution. The quoted errors in \scn{vreal} and \scn{vimag} also
include effect of varying the fit range
within the plateau region, and spread over smearing levels (see
\scn{system}).  

We present results for the real part of the potential in \scn{vreal},
and in \scn{vimag} the results for the imaginary part are
shown. Discussion of various systematics related to the results
presented in this section have been put in \apx{system}.

\bef
\centerline{\includegraphics[width=7cm]{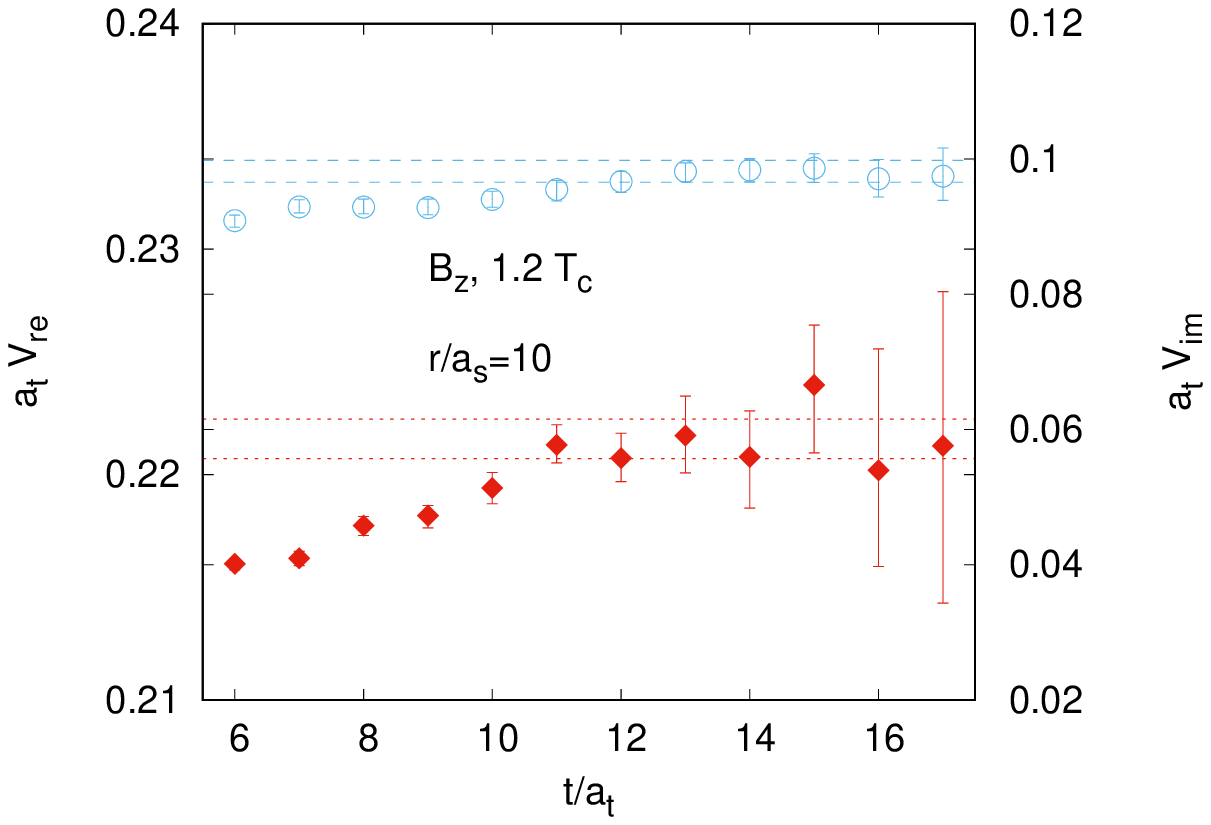}
  \includegraphics[width=7cm]{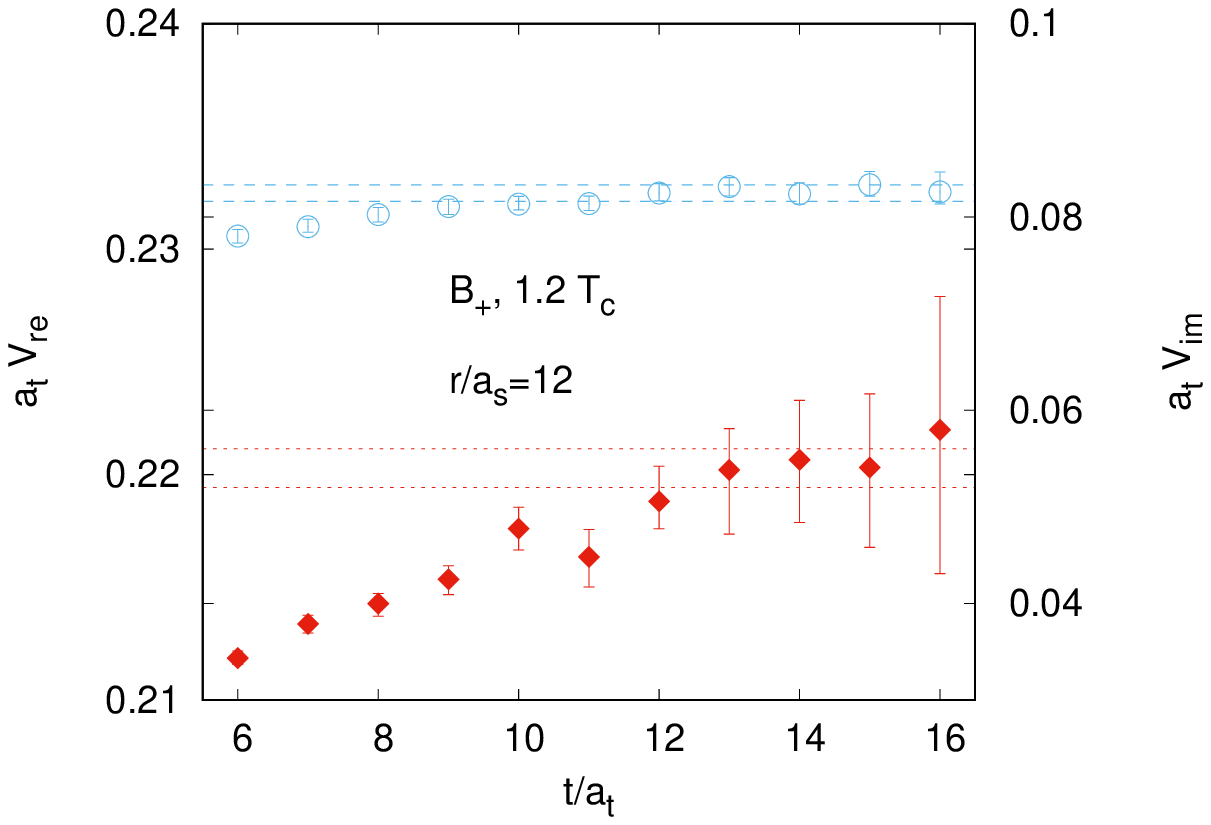}}
\caption{``Local values'' of the potential, $\vro$ and $\vim$, for  
  Set 3, at 1.2 $T_c$. Results for smearing level 300 are shown.
  The results quoted in \scn{vreal} and \scn{vimag}, obtained from
  fits over the plateau range, are shown with horizontal bands.  The
  filled diamonds and dotted lines show results for $\vim$ while the
  empty circles and dashed lines show results for $\vre$. (Left panel)
  results for L=0, at $r=10 \, a_s$. (Right) Those for L=1, at $r=12
  \, a_s$.}
\eef{plateau}

\subsection{$\vro$}
\label{sec.vreal}
As discussed in \scn{analysis}, the results for $\vro$ are obtained
from $\wap$ with the hybrid current operators. At $T=0$ the hybrid
potential has been studied in detail in the literature. In \fgn{vro0}
we show the potential obtained by us below $T_c$ for the two operator
insertions $\bz$ and $\bxy$.  Strictly speaking the lattices here are
at 0.75 $\tc$ (see \tbn{sets}); but in gluon plasma one expects very
little temperature effect at this temperature, and we indeed checked
that our results are in very good agreement with a recent analysis of
T=0 hybrid potential, Ref. \cite{octet0}. Since this is, in  effect, a
zero temperature potential, we obtained the potential from a
conventional exponential fit. 

\bef
\centerline{\includegraphics[width=7.5cm]{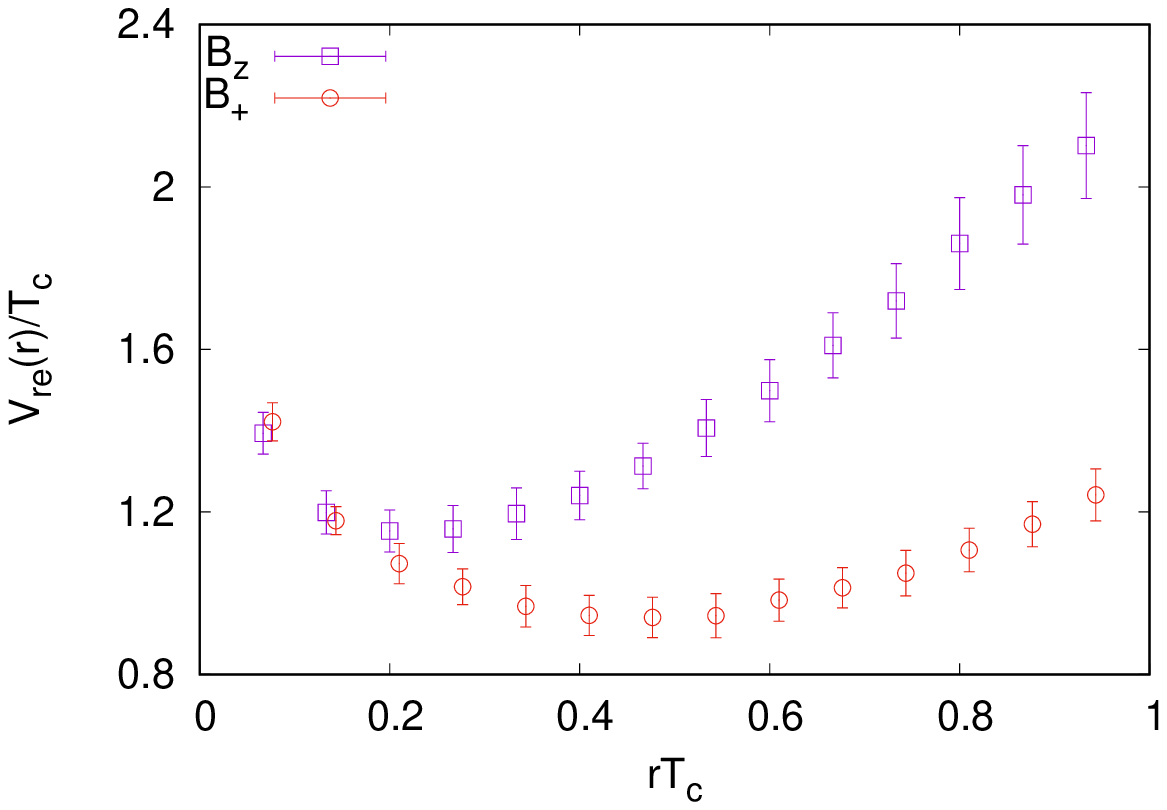}}
\caption{Hybrid potentials below $T_c$ for the L=0 ($\bz$) and L=1
  ($\bxy$) channels. The results are obtained from the $30^3\times 60$
  lattices in \tbn{sets}. The results for L=1 channel have been shifted
  slightly along x-axis in the plot, for ease of viewing.}
\eef{vro0}  

From \eqn{hybridj} we would expect that, at short distance, the potentials
extracted for the two channels would agree and give the octet potential,
modulo an additive constant. As the figure shows, the potentials
do seem to agree at very short distances $\lesssim 0.1$ fm, showing the
repulsive behavior expected of the octet channel.
A detailed analysis of the short distance part has been made in
\cite{balipineda}.
At larger distances, nonperturbative effects start dominating, and the
potentials for the two channels have a different nature, giving the
hybrid potentials. While both channels
are attractive at long distances, supporting bound states for $\Sigma_u^-$
and $\Pi_u$ respectively, quantitatively the long distance attractive part is
very different for the two channels, and also from the long distance
part of the singlet.

At finite temperatures, the long-distance nonperturbative behavior is
suppressed, and one may expect to be able to extract
information about the octet potential over longer distances. This is
exactly what we found for $\vro$. We extract
the potential from $\wap$, as explained in \scn{analysis}. 

In \fgn{vrL0L1} we compare the results of $\vro$ extracted from the
Wilson loop with $\bz$ and $\bxy$ insertions. Note that the potential
can be extracted from the Wilson loop modulo an additive
renormalization constant (see \apx{lo}). For the results in this
section, we have fixed the additive renormalization constant by
demanding that the T=0 singlet potential at the shortest distance
$r=a_s$ matches the lattice discretized Coulomb potential:
\beq
V_s(r=a_s, T=0) \ = \ - g^2 C_F \dtk(k) \ 
\frac{\cos k_3 a_s}{4 \, \sum_i \, \sin^2 (k_i a_s/2)}
\eeq{latcoul}
where for the coupling $g^2$ we have used the ``boosted
coupling'' $g^2(r \sim a) \, = \, \frac{\textstyle 6}{\textstyle \sqrt{\bs
    \, \bt} \; \sqrt{P_s \, P_t}}$; $P_s, P_t$ are the plaquette
variables defined in \eqn{plaq}. The choice of the coupling and the
matching point is somewhat arbitrary, and no detailed study of
systematics of the subtraction was done; so the potentials 
should be taken to be defined modulo a small, temperature-independent
additive constant. We stress that the additive normalization of
the singlet at $T=0$ fixes the renormalization both $\vrs$ and $\vro$
at all temperatures.

The difference between \fgn{vrL0L1} and \fgn{vro0} is stark: there is
no nonperturbative rising part of the potential above $\tc$, and the
potentials for L=0 and L=1 agree very well (within errors)
to the distance studied. Here, therefore, free from any dominant
effects of the gluon string, one can talk about a ``octet'' potential,
which is related to the interaction between the heavy quark and
antiquark in the color octet configuration.  

\bef
\centerline{\includegraphics[width=6.5cm]{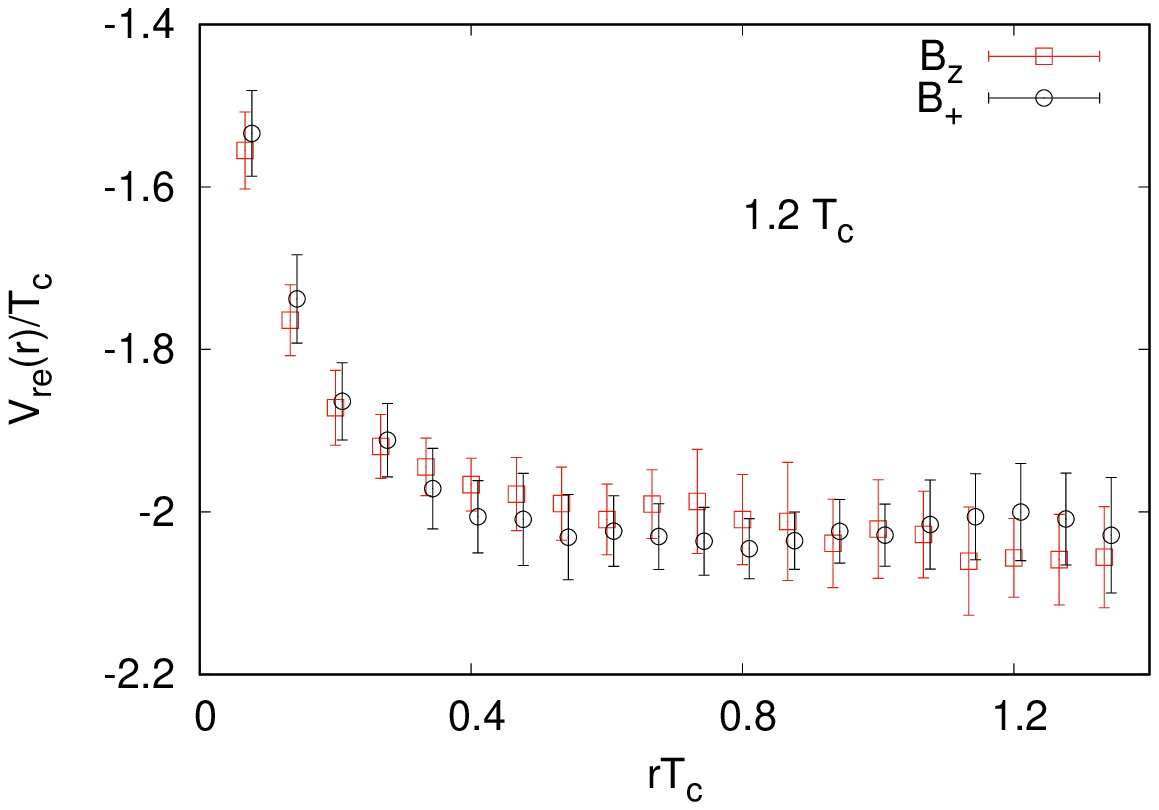}
  \includegraphics[width=6.5cm]{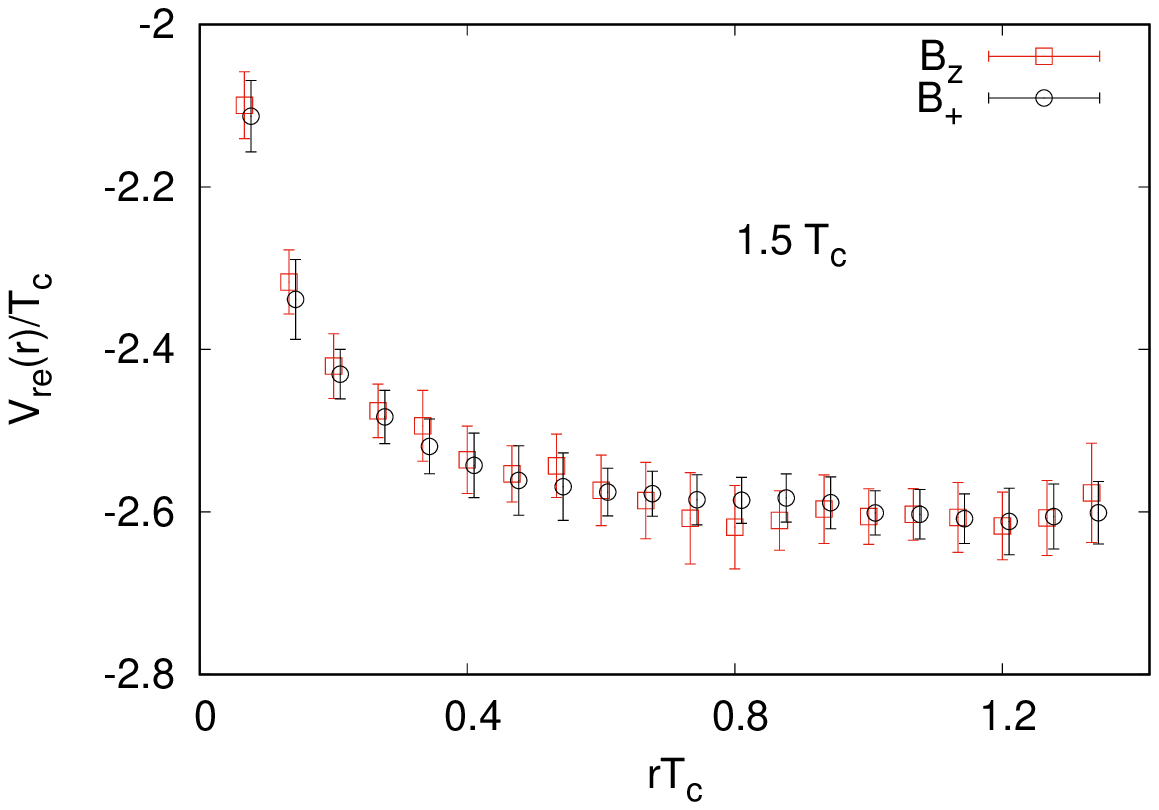}
    \includegraphics[width=6.5cm]{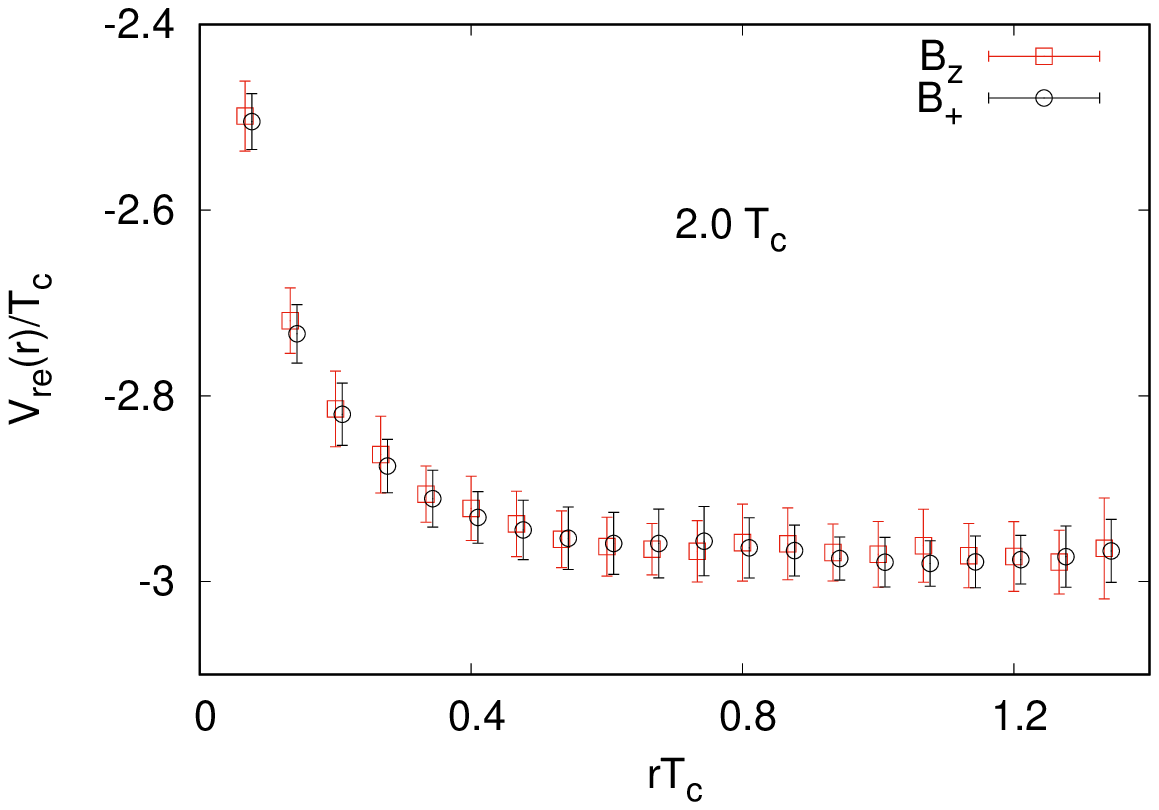}}
\caption{Comparison of the $\vro$ obtained for the L=0 ($\bz$) and L=1
  ($\bxy$) channels. (Left) 1.2 $\tc$, (middle) 1.5 $\tc$ and (right)
  2.0 $\tc$. The points for $\bxy$ have been shifted
  slightly along x-axis in the plot.}
\eef{vrL0L1}

In \fgn{vrTdep} we show $\vro$ at different temperatures above $\tc$. 
For comparison, the hybrid potential of \fgn{vro0} is also shown in
the same plot. A constant has been
subtracted from the hybrid potential below $\tc$ for showing it in the
same scale. This figure clearly displays the effect of the
deconfinement transition on the potential: the octet potential above
$\tc$ is repulsive at all distances. There is no trace of the
long distance nonperturbative attractive part present in the
potential of the hybrid operator below $\tc$.

While this is the first nonperturbative study of $\vro$ from lattice,
a related quantity, the free energy of a color octet $\qqb$ pair in
the plasma, has been nonperturbatively studied before
\cite{free,tumfree}. Unlike the
thermal potential, it is straightforward to nonperturbatively define
the color octet free energy of a $\qqb$ pair in the Coulomb gauge.
The free energy is real, and is identical to the potential at
T=0. In the plasma, it has been found to be close to the real part of
the potential; see \cite{singlet} for a comparison of the two for the
singlet channel. The color octet free energy defined in Coulomb gauge
shows screening, and is qualitatively similar to the behavior of
$\vro$ shown in \fgn{vrTdep}. A detailed analysis of the short
distance behavior of the Coulomb gauge fixed color octet free energy
has been done in Ref. \cite{tumfree}, and has been found to be in
excellent agreement with perturbation theory.

\bef
\centerline{\includegraphics[width=7.5cm]{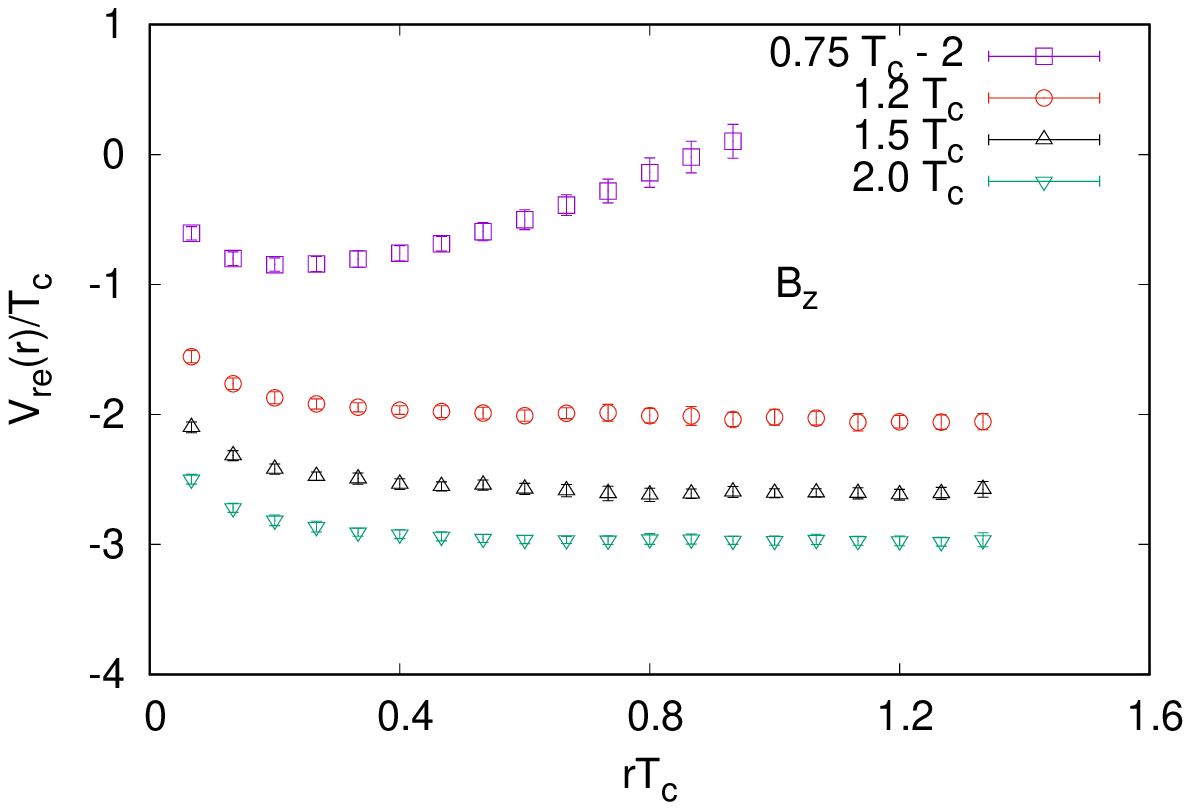}
  \includegraphics[width=7.5cm]{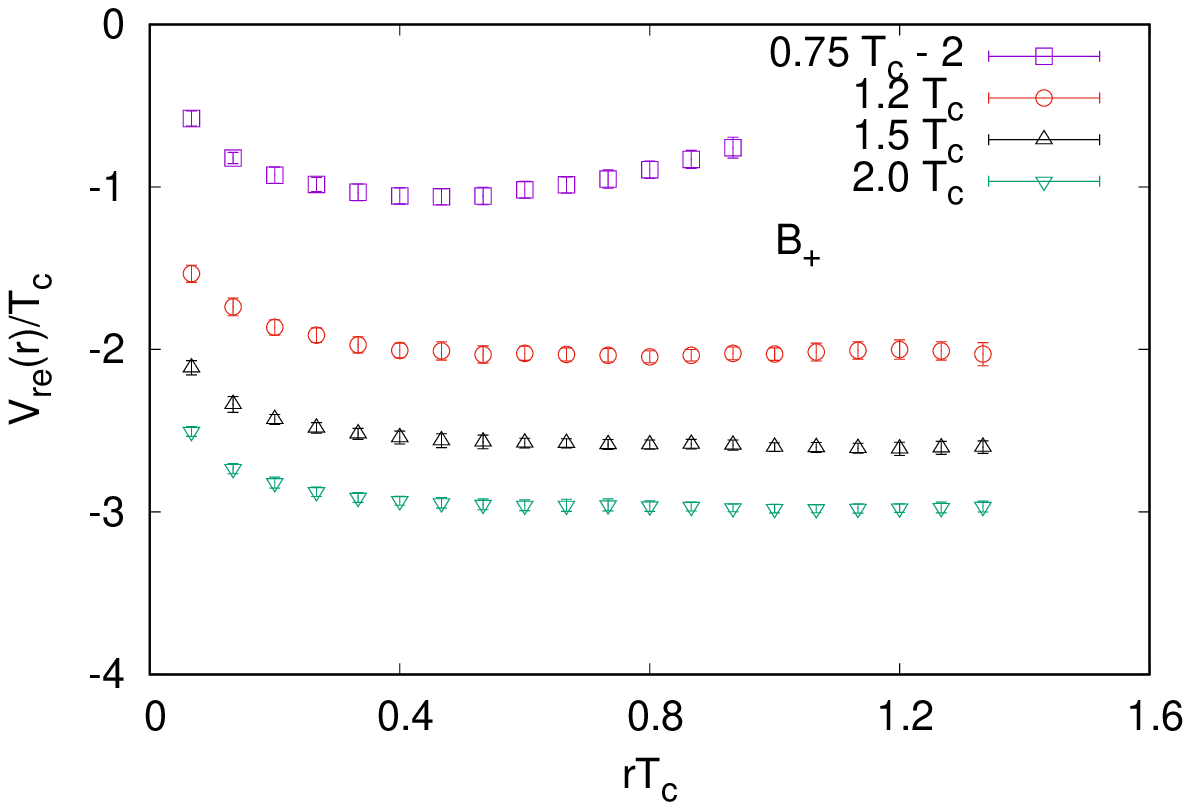}}
\caption{Temperature dependence of $\vro$; (left) L=0 and (right) L=1
  channels. As indicated in the label, a constant has been subtracted from
the potential below $\tc$ for convenience of showing it with the above
$\tc$ results.}
\eef{vrTdep}

In \fgn{vr-so} we display together the octet and singlet potentials
above $\tc$. The octet potential is much flatter
than the singlet potential. Also at each temperature, the two
potentials approach the same temperature-dependent constant.
Physically one expects this; at sufficiently long distance
the interaction between the $Q$ and the $\bar{Q}$ is expected to vanish;
the remnant constant then may be interpreted as a thermal correction
to mass of the quark.
In the right panel of \fgn{vr-so} we check this behavior down to
longer distances using the larger, but coarser, set 1 data. Both
these behaviors are qualitatively consistent with the expectations
from perturbation theory, \eqn{octlo} and \eqn{impots}. 
We also note
that the convergence of the singlet and octet potentials happen at
shorter distances at higher temperatures. This is also expected, since
the difference is $\propto \frac{\textstyle e^{-\md r}}{\textstyle
  r}$, which is smaller at higher temperatures, where $\md$ is larger.
However, there are some quantitative differences from perturbation
theory: in particular, the
differences in the asymptotic values at two temperatures is larger
than what \eqn{octlo} predicts.

\bef
\centerline{\includegraphics[width=7.5cm]{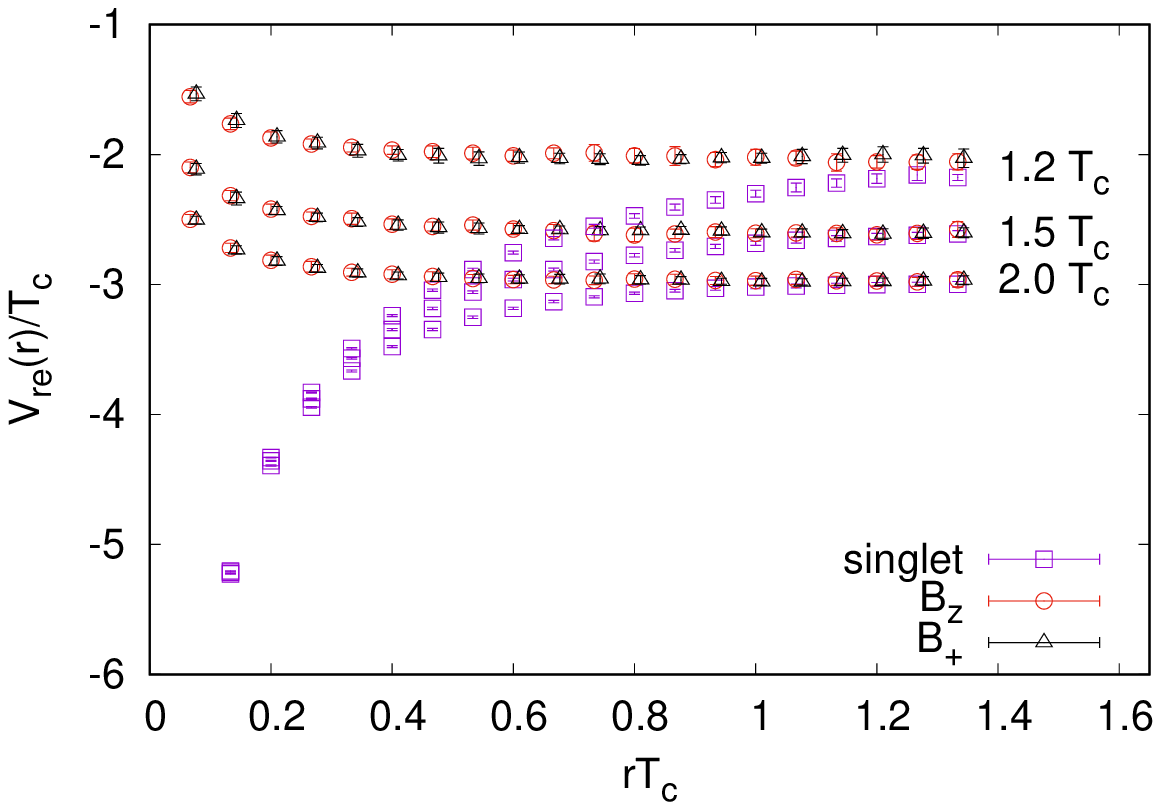}
  \includegraphics[width=7.5cm]{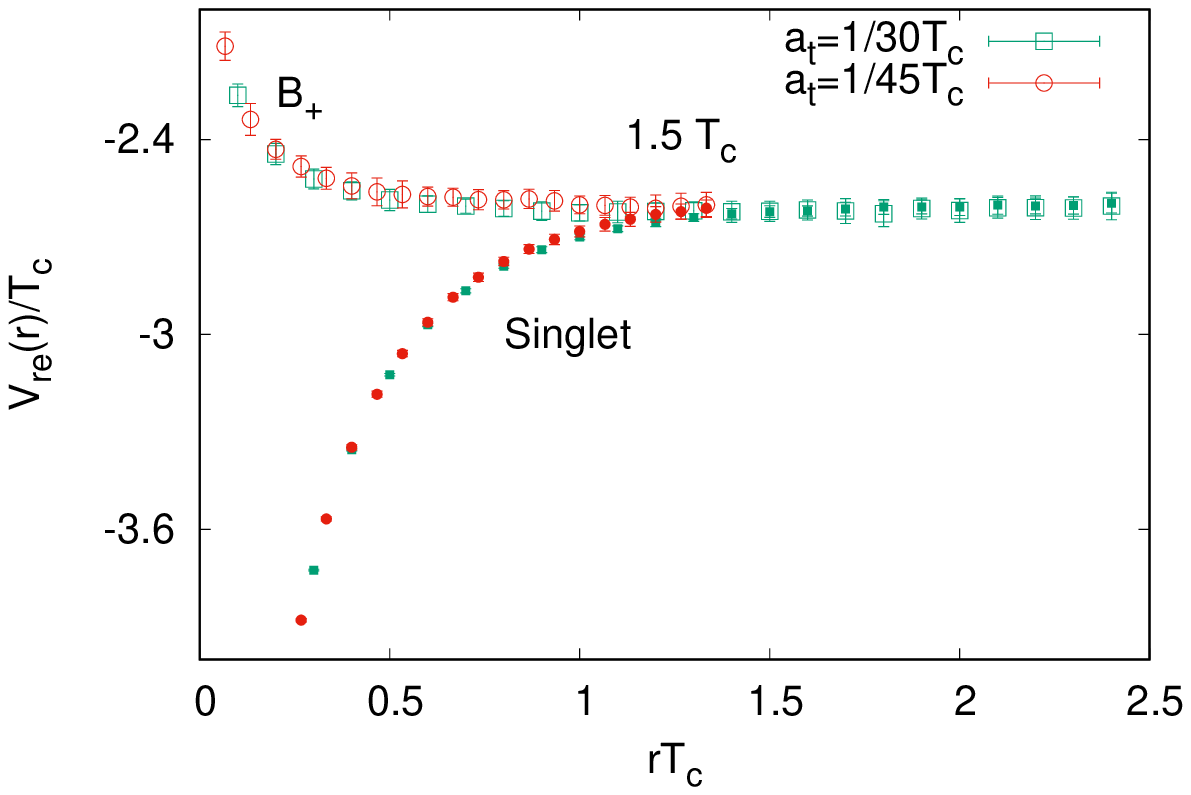}}
\caption{(Left) Comparison of the singlet and octet potentials $\vre$
  above $\tc$.  For $\vro$, results for both L=0 and L=1 channels are
  shown (with the L=1 points slightly shifted along x axis for ease
  of viewing).
  In the right panel, we show this comparison down to a
  distance $r T_c \sim 2.5$, for 1.5 $T_c$. In this plot the long
  distance part is from the coarser lattice of set 1.}
\eef{vr-so}

To further check the conformity with the perturbative behavior, we
look at $\dvr(r) \; = \; \vre(r+1) \, - \, \vre(r)$.
In leading order perturbation theory, the ratio of this quantity
in singlet and octet channels is $-(N_c^2-1)$. As \fgn{dvr} shows, our
data is consistent with this.

\bef
\centerline{\includegraphics[width=6.5cm]{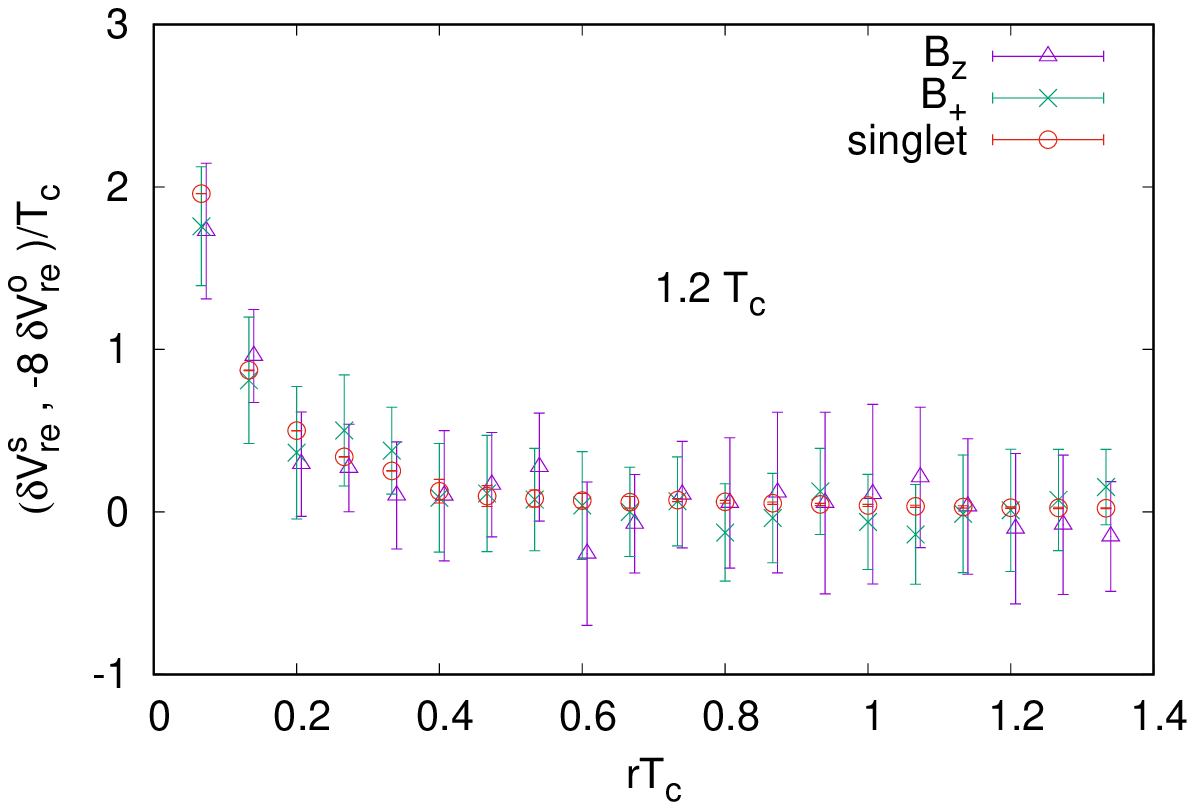}
  \includegraphics[width=6.5cm]{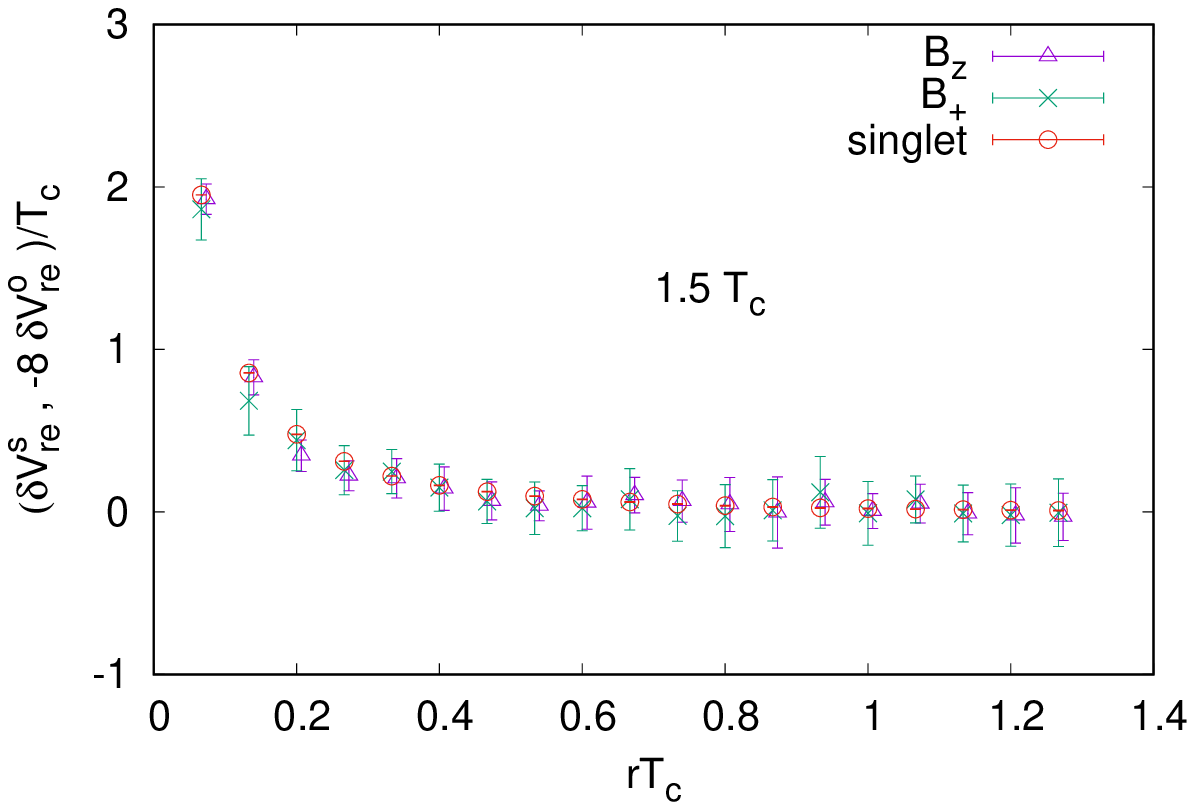}
  \includegraphics[width=6.5cm]{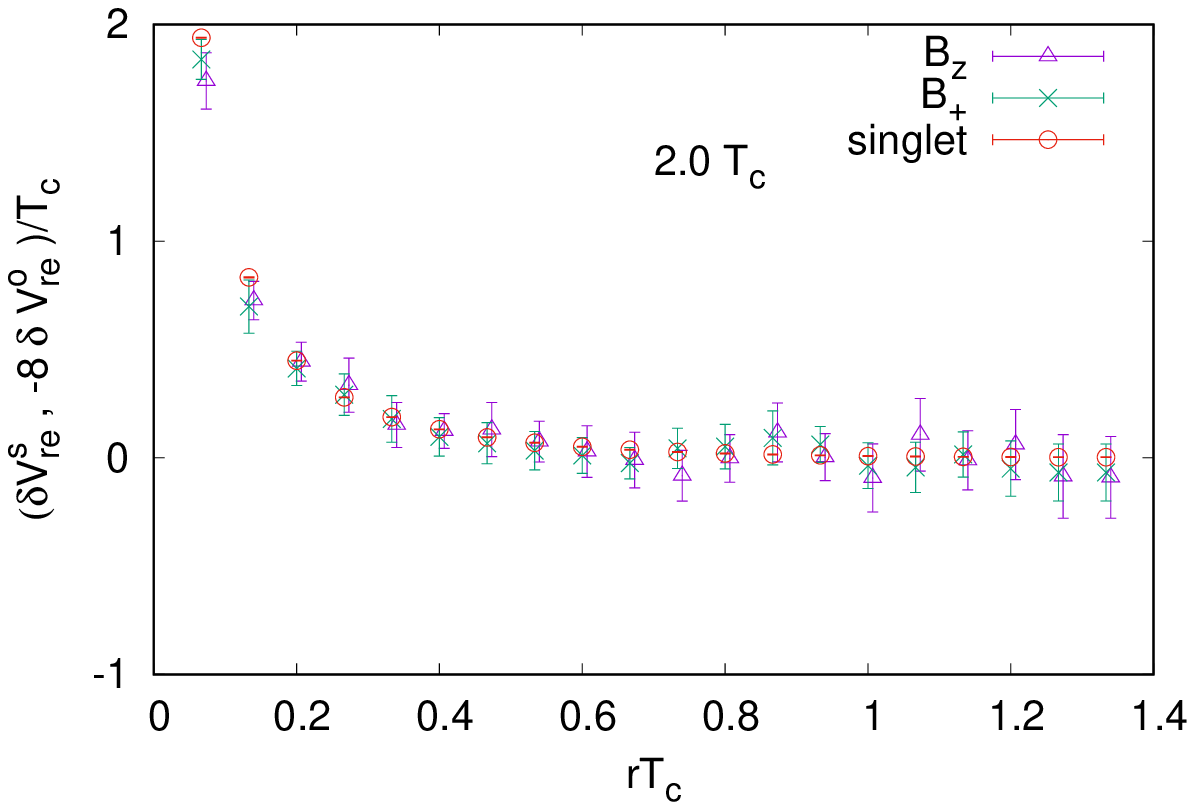}}
\caption{Comparison of $\dvr(r)=\vre(r+1)-\vre(r)$ for singlet and
  octet channels, at 1.2 $\tc$ (left), 1.5 $\tc$ (middle) and
  2 $\tc$ (right). The points for $\bz$ have been slightly shifted along
  the horizontal axis.}
\eef{dvr}

\subsection{$\vio$}
\label{sec.vimag}
The effective thermal potential is in general complex \cite{impot},
with the imaginary part of the potential related to damping and
decoherence mechanisms \cite{aka13}. In the case of singlet
channel, where the potential is attractive and leads to spectral
function peaks for sufficiently massive quarks, the imaginary part
controls the width of the spectral function peak. Such an
interpretation is not available here, and the imaginary part is to be
understood as introducing decoherence in the $\qqb$ system during its
evolution as octet.

The extraction of the imaginary part from the hybrid operators of the
sort used here turns out to be more problematic, and we can only get
partial information about them. In particular, we were not able to get
reliable results for $\vio$ at 2 $\tc$, and show results
only for 1.2 $\tc$ and 1.5 $\tc$. 

In \fgn{viL0L1} we
compare the results for $\vio$ with insertion of the $\bz$ and $\bxy$
operators. Here we see a different behavior from that seen in
\fgn{vrL0L1} for $\vro$: while at short distances, the results for the
two insertions agree within statistical error, at longer distances $rT
\gtrsim 1$ they have statistically significant differences. At
such distances, it is clearly not viable to talk of the extracted
$\vio$ as  color octet potential, as it has contributions from the
gluonic operator insertion.

\bef
\centerline{\includegraphics[width=7.5cm]{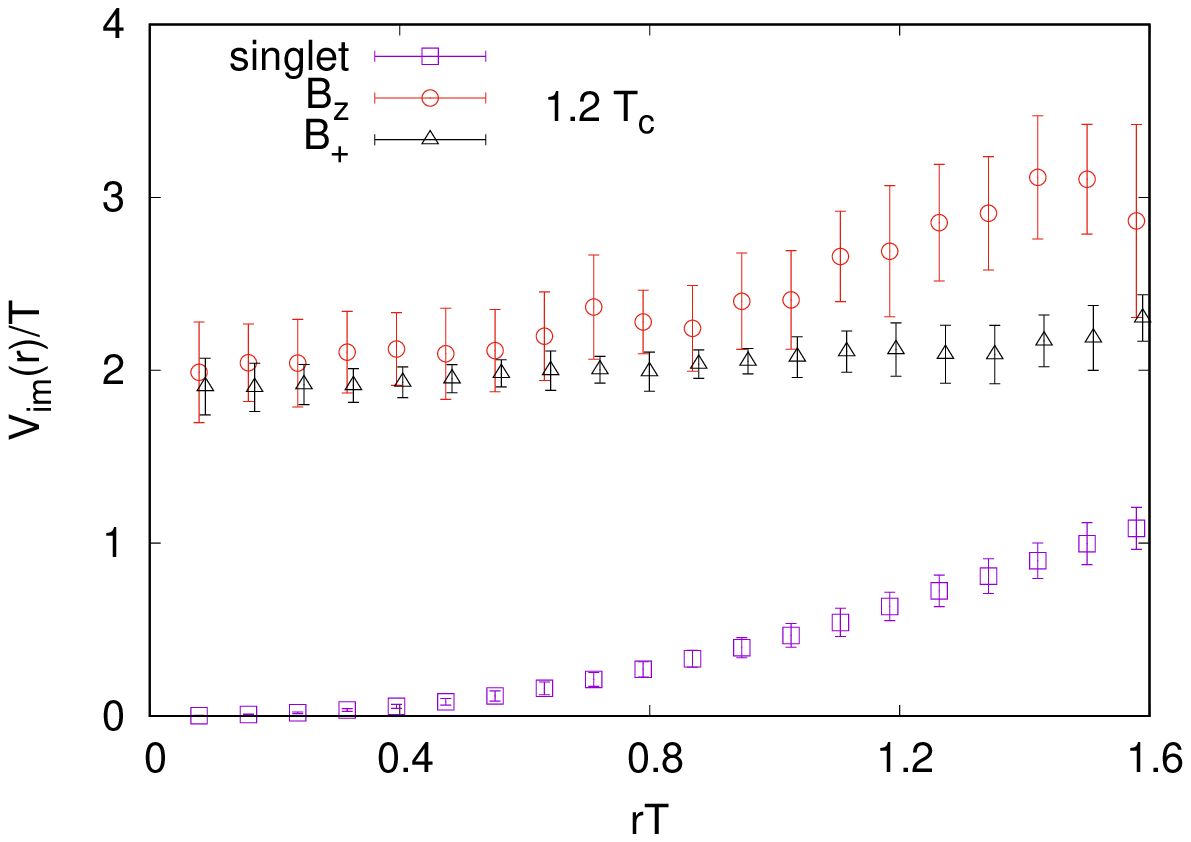}
  \includegraphics[width=7.5cm]{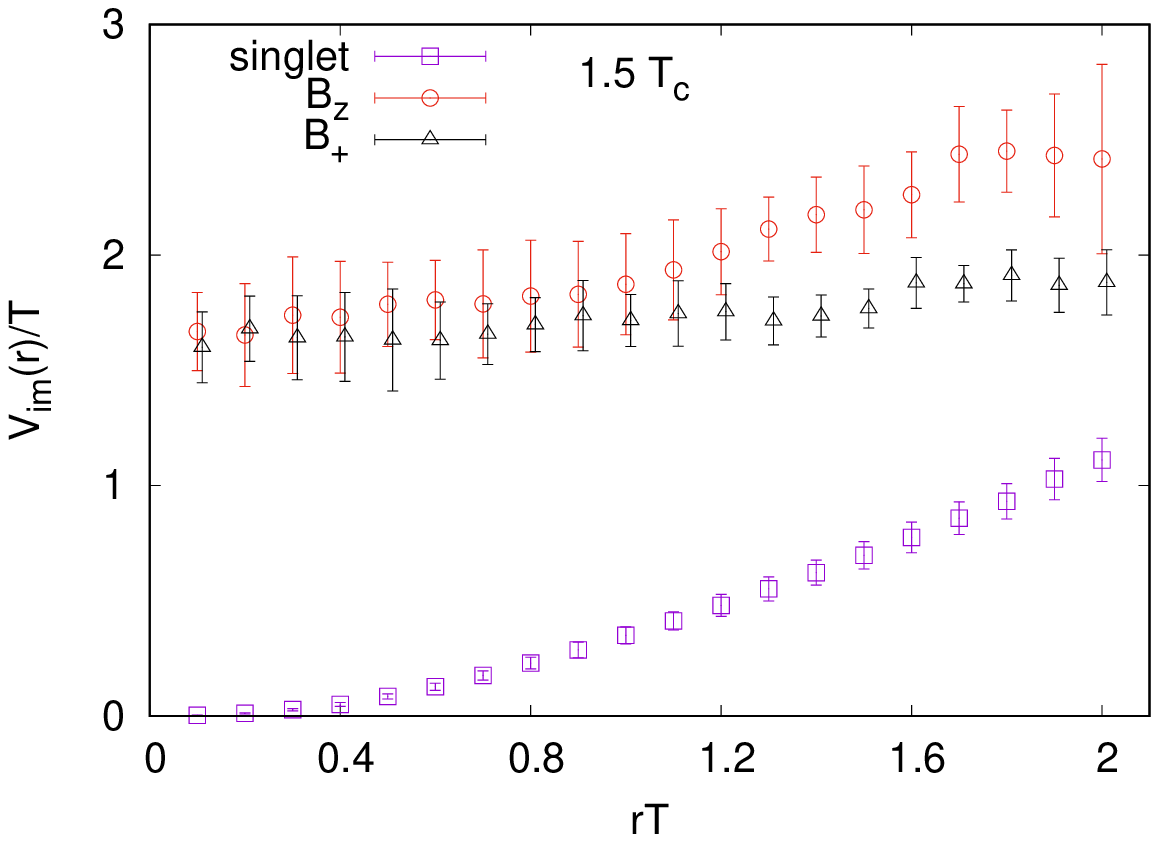}}
\caption{$\vio$ extracted from L=0 ($\bz$) and L=1 ($\bxy$) channels,
  at 1.2 $\tc$ (left) and 1.5 $\tc$ (right). The points for $\bxy$ have
  been slightly shifted horizontally. The singlet channel result at the same
  temperature, $\vis$, is also shown.}
\eef{viL0L1}

In \fgn{viTdep} we show the temperature dependence of $\vio$. Below
$\tc$ $\vio$ is consistent with zero. Above $\tc$ it is very different
from zero. In particular, the most striking behavior of $\vio$ is that
at $r \to 0$ it approaches a finite value. 
It is interesting to see that this behavior is qualitatively
consistent with the behavior suggested in \eqn{octlo}, though of
course, \eqn{octlo} is based on HTL perturbation theory and is not
valid in short distances.

\bef
\centerline{\includegraphics[width=7.5cm]{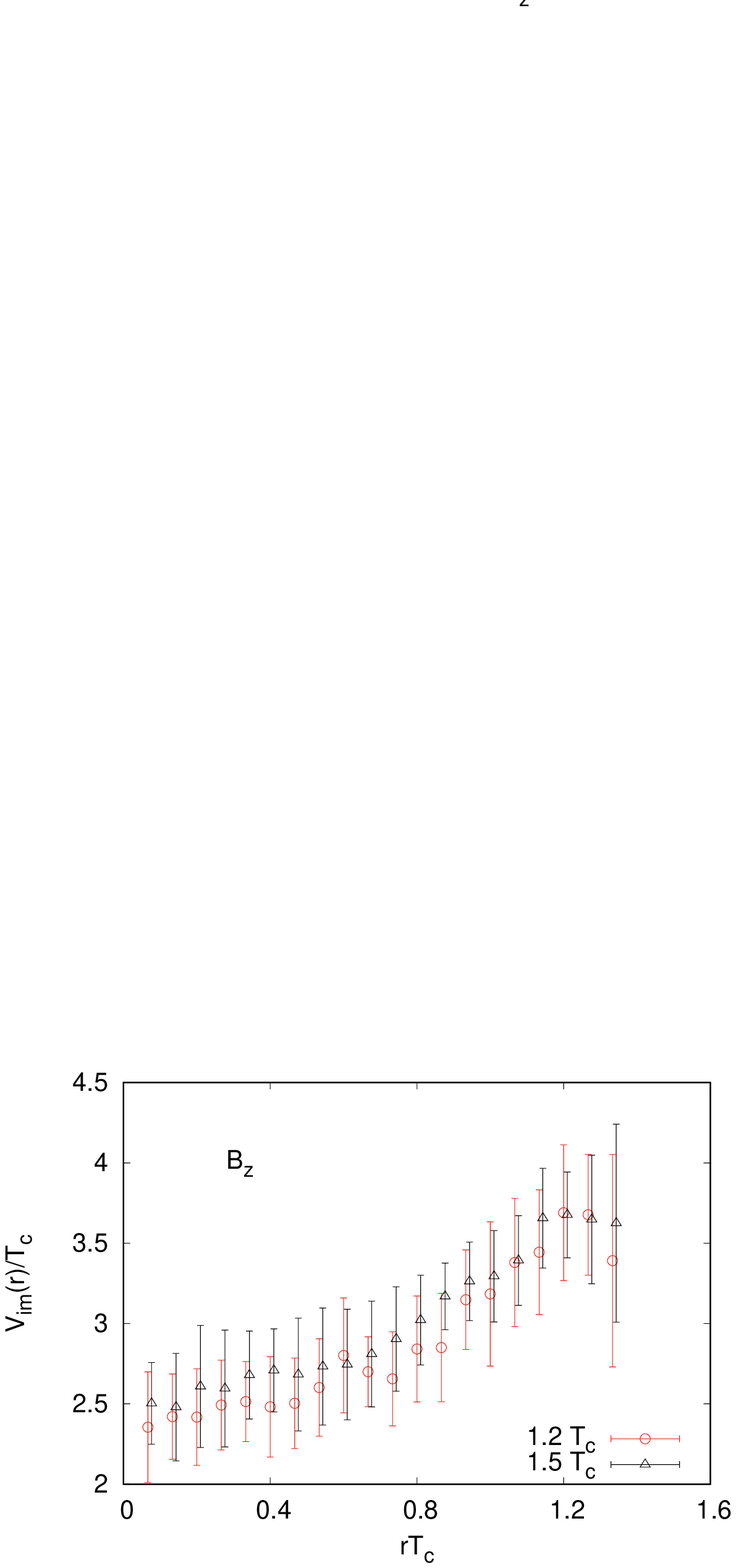}
  \includegraphics[width=7.5cm]{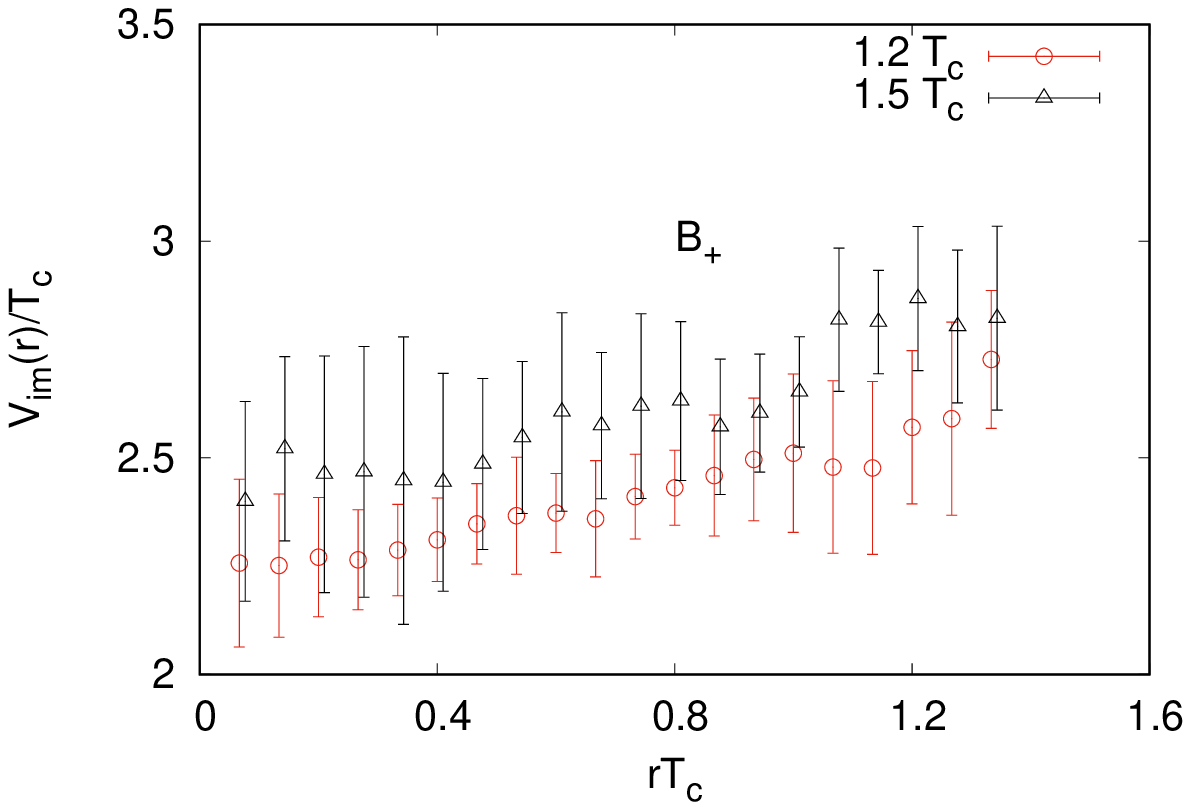}}
\caption{Temperature dependence of $\vio$, for L=0 (left) and L=1
  (right) channels. The points for 1.5 $\tc$ have been slightly shifted
   along the horizontal axis.}
\eef{viTdep}

This behavior is very different from that of the singlet
channel, where $\vis (r \to 0) \to 0$ \cite{singlet}. In \fgn{viL0L1}
we have also shown the imaginary part of the singlet potential.
We find that $\vis$ vanishes at $r \to 0$, but rises $\sim r^2$ at
small $r$ \cite{singlet}.  The slope of $\vio$ at small $r$ 
is smaller than that of $\vis$. Within the accuracy of our results,
$\vio$ for $rT < 1$ is consistent with a constant function.
We do not find evidence of
$\vio$ decreasing with $r$, as suggested by \eqn{octlo}. 

\section{Discussion}
\label{sec.summary}
One way to understand the behavior of a heavy $\qqb$ pair in quark-gluon
plasma is through the introduction of an effective thermal potential
\cite{impot,pnrT,kaar}. In order to understand the evolution of
quarkonia in plasma, we need to know the effective potential of
$\qqb$ pair in both singlet and octet color configurations.

Nonperturbative information about the singlet potential is available
in the literature \cite{singlet}. The real part of the potential shows
the expected medium screening, but at a quantitative level, differs
from the leading order perturbative potential even at 2 $\tc$. The
deviation from perturbation theory is even stronger in the imaginary
part of the potential.

In contrast to the singlet potential, very little is known
nonperturbatively about the in-medium interaction of $\qqb$ in a color
octet configuration. One reason for this is the difficulty in
nonperturbatively defining a color octet potential. In this paper, we
have made the first nonperturbative study of the effective interaction
potential for $\qqb$ in color octet configuration in the plasma. 
The color-octet state is studied by looking at gauge
invariant states formed by combining gluonic operators with color
octet static $\qqb$ source. At $T=0$ potentials for such states,
called hybrid states, have been studied in detail: in the perturbative
regime at small $r$, they are expected to give information about the
octet potential, while at longer distances, where nonperturbative
effects dominate, the potential becomes dependent on details of the
gluonic operator.

In contrast, we find that the in-medium potential $\vro$ above $\tc$
remains independent of the specific hybrid channel, and gives
information about the interaction between the octet $\qqb$ pair. This is
illustrated in \fgn{vrL0L1}. Our results for the color octet potential
is summarized in \fgn{vrTdep}. The color octet potential is found to
be screened above $\tc$, and is repulsive at all distances. This indicates that
there will not be any bound states of the heavy $\qqb$ in the plasma
in the color octet configuration. At long distances, the potential agrees
with the color singlet potential (\fgn{vr-so}). Within the accuracy of
our data, the data is also consistent with the leading order scaling behavior
$\frac{\textstyle \delta \vro}{\textstyle \delta \vrs} \; =
\; - \frac{\textstyle C_A/2 - C_F}{\textstyle C_F}$; this is
demonstrated in \fgn{dvr}.

The thermal effective potential is known to have an imaginary part
\cite{impot}. The imaginary part is related to the physics of Landau
damping and decoherence of the wave function of the $\qqb$ state.
The effective potential obtained from the hybrid state also has an
imaginary component. The extraction of this part is more difficult,
and our results for $\vim$ have large errors. The
imaginary part of the extracted potential is more difficult to
interpret in terms of a color octet potential. For the two hybrid
operators we looked at, we found
agreement in $\vio$ only up to  distances $r T \sim 1$; see \fgn{viL0L1}.
The details of the gluonic operator become important at larger distances,
and one cannot meaningfully identify the extracted potential as octet
potential beyond such distances.

The most striking difference between the imaginary parts of the
singlet and the octet potential, shown in \fgn{viL0L1}, is the behavior
at short distances. $\vis$ approaches zero at short distances. On the
other hand, $\vio$ acquires a nonzero value even at $r \to 0$ on crossing
$\tc$. This is consistent with the behavior predicted in perturbation theory,
and is also in line with physical intuition \cite{aka13}. The singlet $\qqb$ at
very short distances will look like a colorless object to the medium particles, 
which will not be able to resolve its structure. On the other hand,
the medium particles will interact strongly with the color octet $\qqb$,
leading to damping. The $r$ dependence of the imaginary part is much milder
than that of the singlet in the region $rT \lesssim 1$; within the (limited)
accuracy of our calculation, $\vio$ is consistent with a constant
in this region. 

{\bf Acknowledgements:} We would like to thank Gunnar Bali, Nora
Brambilla, Peter Petreczky, Anurag Tiwari and Antonio Vaio for
discussions. This work was carried out under the umbrella of ILGTI.
The computations reported here were performed on the clusters of the
Department of Theoretical Physics, TIFR. We would like to thank Ajay
Salve and Kapil Ghadiali for technical support.

\appendix
\section{APE smearing}
\label{sec.ape}
As discussed in \scn{lat}, the extraction of the potential from thin
link Wilson loops is difficult, and we do APE smearing \cite{ape}
of the spatial gauge links. This consists of replacing the spatial
gauge links
\begin{eqnarray}
U_i (\vec{x}, \tau) \ \rightarrow \ {\rm Proj}_{SU(3)} &{}& \Bigl[ 
\alpha \, U_i(\vec{x}, \tau) + \sum_{\substack{1
    \le j \le 3 \\ j \ne i}} 
\left\{ U_j(\vec{x}, \tau) \,  U_i(\vec{x}+a_s \hat{j}, \tau)  \, 
U_j^\dagger(\vec{x}+a_s \hat{i}, \tau) \right. \label{ape} \\
&{}& \qquad + \, \left. U_j^\dagger(\vec{x}-a_s \hat{j}, \tau)  \,
U_i(\vec{x}-a_s \hat{j}, \tau) 
\, U_j(\vec{x}-a_s \hat{j}+a_s \hat{i}, \tau) \right\} \Bigr] \nonumber
\end{eqnarray}
iteratively. While the quality of the signal for the Wilson loop
detoriates with the number of smearing steps, the effect of
non-potential terms also decrease, making extraction of potential
easier. For this work, we have taken $\alpha$ = 2.5, and have
done up to 400 steps of APE smearing. 

It is instructive to see the effect of APE smearing on the leading
order expressions for the Wilson loops. Following \cite{bernard} we
write the effect of smearing on the gauge fields, $A_\mu$, where
$V(x, x+a \mu) \; = \; e^{\textstyle i \, a \, A_\mu(x)}$. To linear order,
\beq
A_i^N(Q) \ = \ \left\{ f^N(\hat{\vec{q}}) \; P^T_{ij}(\hat{q}) \ + \
P^L_{ij}(\hat{q}) \right\} A_j(Q) 
\eeq{Asmr}
where
\[ f(\hat{q}) \; = \; (1-\frac{c}{4} \hat{q}^2) \; \sim \; e^{- \frac{c}{4} \hat{q}^2},
\qquad \hat{q}^2 = \sum_{i=1}^3 \hat{q}_i^2, \qquad \hat{q}_i=2 \sin q_i
a_s/2, \]
$c=\frac{\textstyle 4}{\textstyle 4+\alpha}$, and the projection operators are
defined above. For small $q_i a_s, \hat{q}_i \to q_i a_s$ and
the projection operators become $P^{T,L}_{ij}(q)$ .

Then the propagator of the smeared fields,
\beq
G^N_{ij}(Q) \; = \; \langle A_i^N(Q) \, A_j^N(-Q) \rangle \; \sim \; f^{2N}(q)
\frac{P^T_{ij} (q)}{Q^2+\pit(q,q)}
\eeq{intsmr}
leads to the spectral function representation
\beq
G^N_{ij}(q) \; \equiv \; \intwh(q) 
\frac{\tilde{\rho}_T(q_0,\vec{q})}{q_0-i \omega_k} \, , \qquad
  \tilde{\rho}_T(q_0,\vec{q}) \; \sim \; f^{2N}(q) \, \rhot(q) \, \cdot
\eeq{spectsmear}
This results in a suppression of the nonpotential contribution
to Wilson loop, as we discuss in the next section.

\section{LO calculation of potential in HTL}
\label{sec.lo}
Various strategies in our nonperturbative calculation of the potential
has been motivated by insights from perturbation theory and in
particular, the expression for the Wilson loop in LO HTL
approximation. Here we put together the leading order results for
the thin Wilson loop, \ref{wm} and \ref{wmg}, in this approximation.
This section follows Ref. \cite{impot}.

We use the Coulomb gauge. Then the gluon propagators are:
\beq
D_{00} \, (\omega_n, \vec{k}) \ = \ \frac{1}{K^2 \, + \, \Pi_E(K)} \;
\frac{K^2}{\vec{k}^2} \, , \qquad 
  D_{ij} \, (\omega_n, \vec{k}) \ = \ \frac{1}{K^2 \, + \, \Pi_T(K)} \
\left( \delta_{ij} \; - \;
\frac{k_i k_j}{\vec{k}^2} \right) \; \cdot    
\eeq{cgaugeprop}    
Here $K$ refers to the Euclidean four-momenta $(\omega_n, \vec{k})$.
The spectral functions $\rhoe(k), \rhot(k)$, introduced through
the integral relations
\beq
\frac{1}{K^2 + \Pi_{T,E}(K)} \ = \ \intwh(k) \frac{\rho_{\scriptscriptstyle T,E} (k_0,
  \vec{k})}{k_0 \, - \, i \omega_k} \, ,
\eeq{spectraldef}
provide the connection to Minkowski momenta.

For the singlet channel, the potential in LO will come from
diagrams for ordinary Wilson loop similar to the ones shown
within parentheses of \eqn{factor}. They add up to
\beq
g^2 C_F \dtk(k) \left( \cos k_3 r \, - \, 1 \right) \ \left\{
\frac{\tau}{\vec{k}^2 \, + \Pi_E(0, \vec{k})} 
+ \intwh(k) \; \rhoe(k) \; (1 \, + \, \bose(k)) \;
\left(\frac{1}{\vec{k}^2} \, - \, \frac{1}{k_0^2} \right)
\; (1 + e^{-\beta k_0} - \ffk(k_0)) \right\}
\eeq{spot}

where we define the symmetric and antisymmetric functions
\beq
\ffk(k_0) \ = \ e^{-k_0 \tau} \; + \; e^{-(\beta - \tau) k_0} \, ,
\ \ \ggk(k_0) \ = \ e^{-k_0 \tau} \; - \; e^{-(\beta - \tau) k_0} \, .
\eeq{sym}      

The term linear in $\tau$ in \eqn{spot} survives in $\wap$, leading to
the potential
\beq
V_r \equiv g^2 C_F \dtk(k) \
\frac{\cos k_3 r \, - \, 1}{\vec{k}^2 \, + \Pi_E(0, \vec{k})} \ \ = \
- \frac{g^2 C_F}{4 \pi r} \; e^{-\md r} \ + \
g^2 \, C_F \left(- \frac{\md}{4 \pi} \; + \; I_a \right)
\eeq{vrtot}
where the additive divergent term $I_a = \dtk(k) \, \frac{1}{\vec{k}^2}
\sim \frac{1}{a}$ in \eqn{vrtot} results from defining the potential
through Wilson loop, forcing $V_r(r \to 0) \to 0$. The standard
convention of defining potential, used in \eqn{impots}, sets
$I_a \to 0$, so that we get the
familiar Coulomb potential at short distances. This has been done in
\scn{vpot} by fixing the T=0 singlet potential at $r=a_s$ through
\eqn{latcoul}.

Going to Minskowski time and taking large $t$, using the relation 
\beq
\lim_{t \to \infty} i \partial_t \ffk(k_0) \vert_{\tau \to it} =
\lim_{t \to \infty}  k_0 \ \left( e^{-i k_0 t} \, - \,
e^{-\beta k_0} e^{i k_0 t} \right) \to - k_0^2 \, 2 \pi i \delta(k_0)
\, \cdot \eeq{limit}
we see that the potential picks up contribution from $\rhoe(k_0 \to
0)$. In leading order of HTL perturbation theory, for $|k_0| \ll
|\vec{k}|$, $\rhot(k), \rhoe(k)$ in \eqn{spectraldef} behave like \cite{impot}
\beq
\rhoe(k) \ \approx \  - \pi \md^2 \frac{k_0}{2 \lvert \vec{k} \rvert
  \; \left(k_0^2 \, + \, \md^2 \right)^2}, \qquad \qquad
  \rhot(k) \ \approx \   \pi \md^2 \frac{\om}{4 \lvert \vec{q}
    \rvert^5}.
\eeq{pertspectral}
The term with $1/k_0^2$ in the second term of \eqn{spot} then leads to
$\vim$ in \eqn{impots}. The $1/k_3^2$ term does not lead to a
potential; \fgn{combo} indicates that the contribution of this
term is small near $\tau=\beta/2$. As discussed in \scn{ape}, smearing
will suppress $\rhot(k)$.

The diagrams
\beq
  \raisebox{-19pt}{
    \begin{axopicture}(60,60)
      \EBox(0,0)(50,50)
      \Gluon(25,0)(25,50){4}{5}
    \end{axopicture}} \
  \raisebox{-19pt}{
    \begin{axopicture}(60,60)
      \EBox(0,0)(50,50)
      \GluonArc(25,0)(10,0,180){3}{4}
    \end{axopicture}} \
  \raisebox{-19pt}{
    \begin{axopicture}(60,60)
      \EBox(0,0)(50,50)
      \GluonArc(25,50)(10,180,0){3}{4}
  \end{axopicture}}
\eeq{nonpot}
add up to
\beq
g^2 C_F \dtk(k) \left( \cos k_3 r \, - \, 1 \right) \; \left(
\frac{1}{k_3^2} - \frac{1}{\vec{k}^2} \right) \intwh(k) \;
\rhot(k) \, (1+\bose(k_0)) \; (1 + e^{-\beta k_0} - \ffk(k_0) ). 
\eeq{snonpot}
Here the gluon lines correspond to transverse gluon propagators. 
\eqn{snonpot} does not contribute to the potential, as can be seen using
\eqn{limit}. However, they will contribute to the fit near
$\tau \sim \beta/2$. These terms, however, have $\rhot(k)$; as
explained in \scn{ape}, smearing leads to a strong suppression of these
terms. When the results for potential stabilize with number of smearing
steps, it indicates that the contribution of these terms have become
negligible and we are getting contribution from the potential terms
only.

The discussion for the hybrid Wilson loop is similar. The potential
contributions in LO come from the diagrams explicitly shown in the
rhs of \eqn{factor}, summing up to

\begin{eqnarray}
\langle B B \rangle \; &{\times}& \left\{ 1 \, + \, \dtk(k) \, 
\left( g^2 C_F \, + \, \frac{g^2}{2 N_c} \, e^{i k_3 r} \right)
\left[ - \frac{\tau}{\vec{k}^2 \, + \, \Pi_E(0, \vec{k})} \right. \right.
  \; \label{octpotexpr} \\
  &{+}&  \left. \left. \intwh(k) \; \rhoe(k) \, \left(\frac{1}{k_0^2}
\, - \, \frac{1}{\vec{k}^2} \right) \, (1+\bose(k_0)) \,
  \left(1+e^{-\beta k_0} - \ffk(k_0) \right) \right] \right\} \nonumber
\end{eqnarray}
Renormalizing in the same way as the singlet leads to the potentials
\eqn{octlo}. We reiterate that the additive renormalization
we have used is fixed by matching of the T=0 singlet potential at
$r/a_s$ = 1: no separate additive renormalization is used for the
octet. 

Other diagrams included in $\cdot \cdot \cdot$ in \eqn{factor} are \\
\beq
  \raisebox{-19pt}{
    \begin{axopicture}(60,60)
      \EBox(0,0)(50,50)
      \GCirc(25,0){1.5}{1.0}
      \GCirc(25,50){1.5}{1.0}
      \GluonArc(10,50)(5,180,0){2}{2}
    \end{axopicture}} \
  \raisebox{-19pt}{
    \begin{axopicture}(60,60)
      \EBox(0,0)(50,50)
      \GCirc(25,0){2}{0.80}
      \GCirc(25,50){2}{0.80}
      \GluonArc(25,50)(9,180,0){3}{4}
    \end{axopicture}} \
  \raisebox{-19pt}{
    \begin{axopicture}(60,60)
      \EBox(0,0)(50,50)
      \GCirc(25,0){2}{0.80}
      \GCirc(25,50){2}{0.80}
      \Gluon(15,0)(15,50){4}{6}
    \end{axopicture}} \
  \raisebox{-19pt}{
    \begin{axopicture}(60,60)
      \EBox(0,0)(50,50)
      \GCirc(25,0){2}{0.80}
      \GCirc(25,50){2}{0.80}
      \Gluon(35,0)(15,50){4}{5}
  \end{axopicture}}
\eeq{octnonpot}

and variations: where the gluon lines are at $\tau=0$ or to the right of
$B$, etc. The sum of their contributions is
\bea
\langle BB \rangle & \times & \left\{ \frac{g^2}{2 N_c} \, \dtk(k) \;
  \frac{1 \, - \, \cos k_3 r}{k_3^2} \; - \; \frac{g^2 N_c}{2} \, \dtk(k)
  \; \frac{2 \, - \, \cos k_3 x \, - \, \cos k_3(r-x)}{k_3^2} \right\}
  \nonumber \\
  &{}& \intwh(k) \, \rhot(k) \; (1+\bose(k_0)) \; \left( 1 \, + \, e^{-\beta k_0}
  \, - \ffk(k) \right) \, \cdot
  \eea{doctnonpot}
Using \eqn{limit} we see that they do not contribute to the
potential. In the Euclidean time data, smearing suppresses
their contribution, due to the $\rhot(k)$ terms.

Diagrams that do not satisfy the factorization behavior of
\eqn{factor} are
\beq
  \raisebox{-19pt}{
    \begin{axopicture}(70,70)
      \EBox(0,0)(60,60)
      \GCirc(30,0){4}{0.80}
      \GCirc(30,60){4}{0.80}
      \Gluon(30,0)(30,30){3}{4}
      \Gluon(30,60)(30,30){3}{4}
      \Gluon(15,60)(30,30){3}{4}
    \Vertex(30,30){2}
  \end{axopicture}} \ + \ {\rm Variations} \ \ + \ \ 
  \raisebox{-19pt}{
    \begin{axopicture}(70,70)
      \EBox(0,0)(60,60)
      \GCirc(30,0){4}{0.80}
      \GCirc(30,60){4}{0.80}
      \Gluon(30,0)(30,30){3}{4}
      \Gluon(30,60)(30,30){3}{4}
      \DashLine(0,30)(30,30){4}
      \Vertex(30,30){2}
    \end{axopicture}} \ + \ {\rm Variation}
\eeq{doctnonfact}
The left set involves only the symmetric function $\ffk(k)$ whereas
the right set involves both $\ffk(k)$ and $\ggk(k)$ of \eqn{sym}. The
expressions are straightforward, if unilluminating; it is easy to
check that they do not contribute to the potential. Also they both
involve two or more factors of $\rhot(k)$ and can be suppressed by
smearing. 

The diagrams \eqn{doctnonpot} and \eqn{doctnonfact} do not contribute
to the potential; however, they contribute in the finite $\tau$
Euclidean Wilson loops. For a successful extraction of the potential,
we need to identify a plateau where their contributions are
negligible. As the structures of these terms demonstrate, smearing
lead to their suppression; that is why we get the plateaus
demonstrated in \fgn{plateau}, from where we can extract the potential.

\section{Systematics in potential estimation}
\label{sec.system}
In this section we discuss the effect of smearing on our extracted
potential, and the size of the discretization error in our results. 

\subsection{$\vro$}
\label{sec.checks}
As discussed in the text, for the spatial gauge connections $\wil$ in
\eqn{wmg} we have used APE smeared links. Smearing also
affects the $G$ field.  For the singlet potential, it was noticed that
the extraction of the potential depends on the smearing level to some
extent \cite{singlet}: one gets a better identified plateau, and the
extracted potential seems to change with the smearing level at small
levels of smearing, before stabilising at some level. A similar trend
is seen in the octet case, except the effect is somewhat enhanced, and
one needs to go to higher levels of smearing before the result becomes
insensitive to the smearing level (within our errors). At higher
levels of smearing, a better plateau is obtained; at the same time the
statistical noise increases. In \fgn{vr-smear} we show the potential
extracted from Wilson loops with different levels of APE smearing, for
set 3. We find that the potential saturates
only at 300 smearing steps at this cutoff. For the results quoted, we
have included a systematic error covering the spread between results
from 300 and 400 levels of smearing. For comparison, for the same set,
the singlet potential stabilised by 200 smearing steps. For set 2, we
find that 200 smearing steps is enough to stabilise the potential, and
the sytematic error covers data with 200 and 250 smearing steps. For
\fgn{vr-smear} as well as for other figures shown in cutoff units, we
show the unrenormalized data (i.e., the matching to \eqn{latcoul} is
not done).

\bef
\centerline{\includegraphics[width=6.5cm]{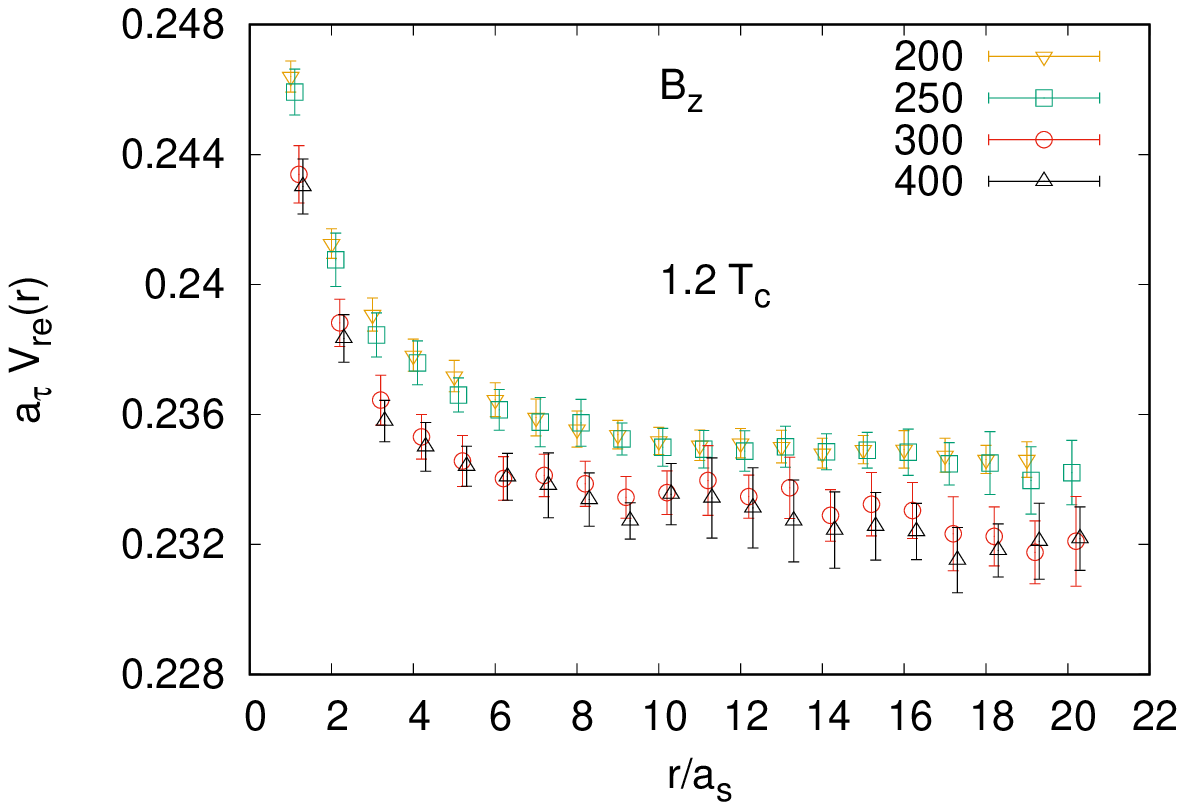}
  \includegraphics[width=6.5cm]{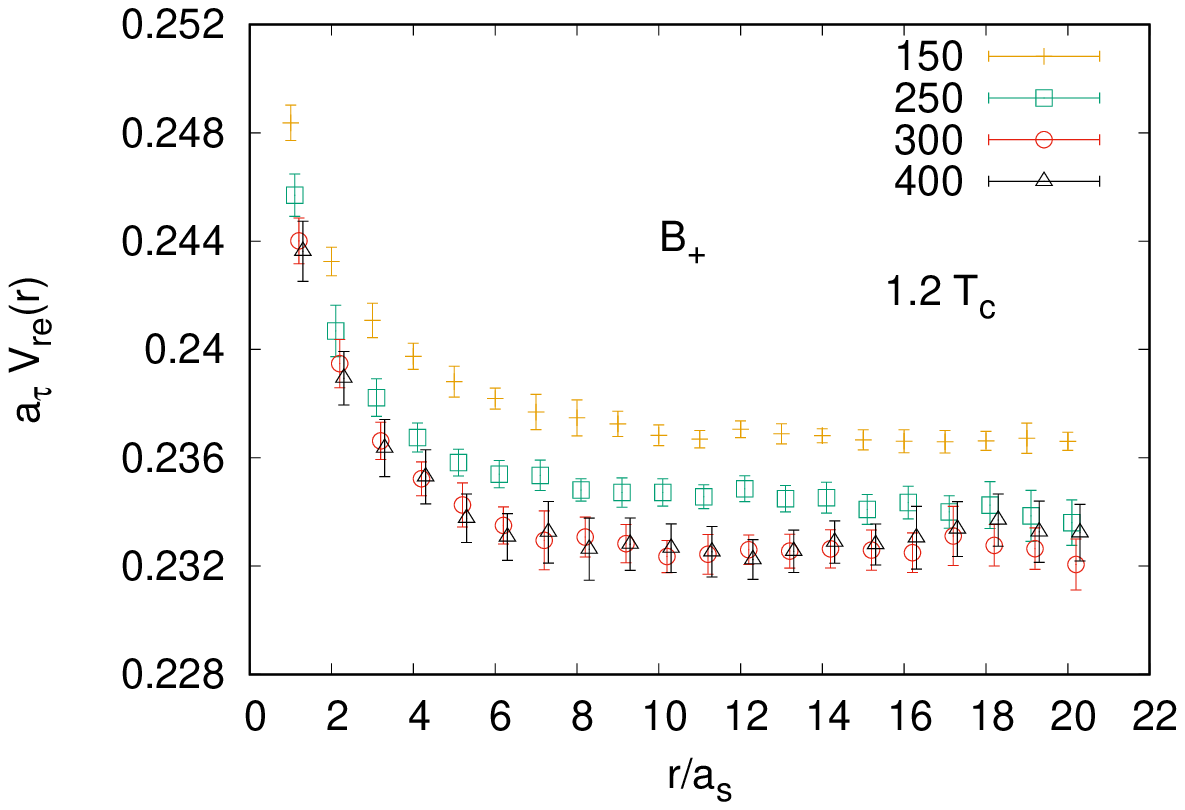}}
\centerline{\includegraphics[width=6.5cm]{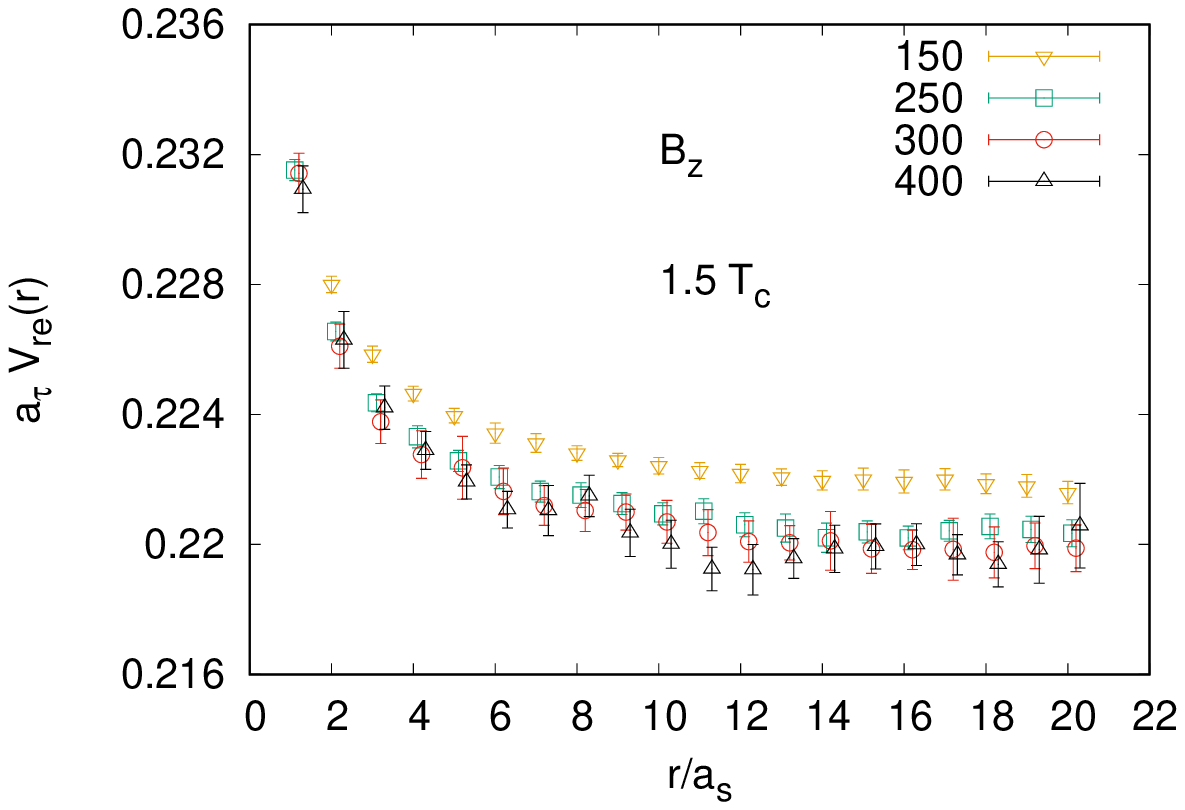}
  \includegraphics[width=6.5cm]{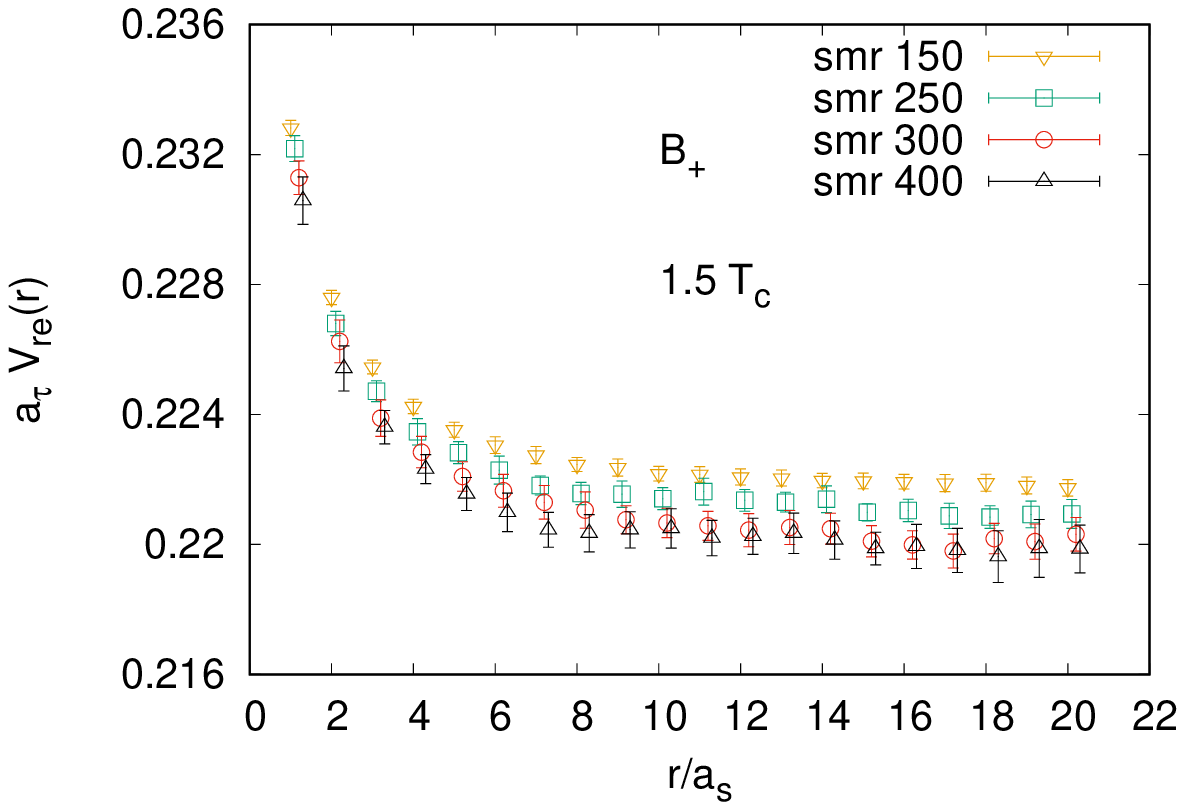}}
\caption{Smearing dependence of our $\vro$ extracted from the
  smeared Wilson loops $\wmg$. The top
  row shows results at 1.2 $\tc$ while the bottom row shows results
  at 1.5 $\tc$. The panels to the left are for $\bz$ and those to
  the right are for $\bxy$ insertions. For ease of viewing, some sets have
  been slightly shifted along x axis in the plot.}
\eef{vr-smear}

The lattice-discretized results will have discretization errors, which
go to $0$ as one takes the continuum limit. Our lattices are quite
fine-grained, so discretization effects are expected to be small.
In order to estimate the size of the discretization error, in
\fgn{vr-cont} we show the results for $\vro$ extracted from two
different lattice spacings. Here we added a small overall additive
constant to the results at the coarser lattice. We see that
the discretization error is much smaller compared to the other
uncertainties in our calculation. We therefore take the results from
our finest set, set 3, as indicative of the continuum results. These
are the results shown in \scn{vreal}. 

\bef
\centerline{\includegraphics[width=6.5cm]{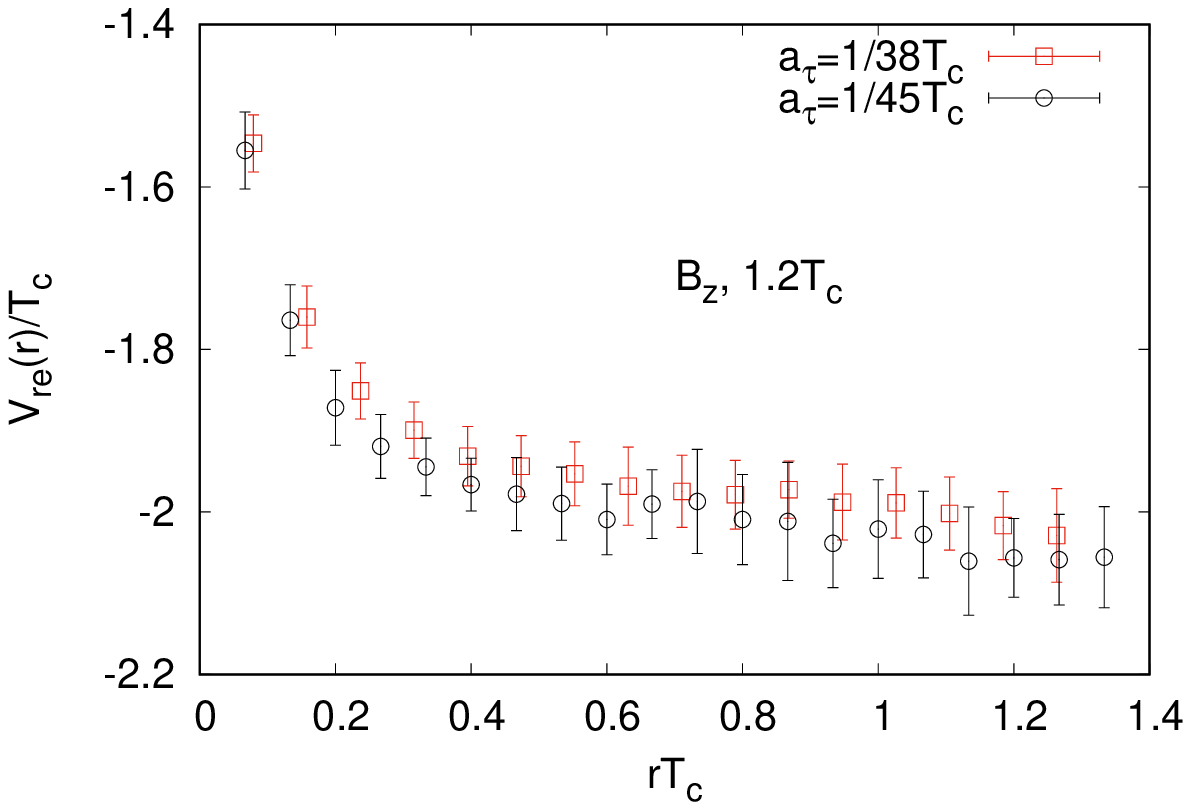}
  \includegraphics[width=6.5cm]{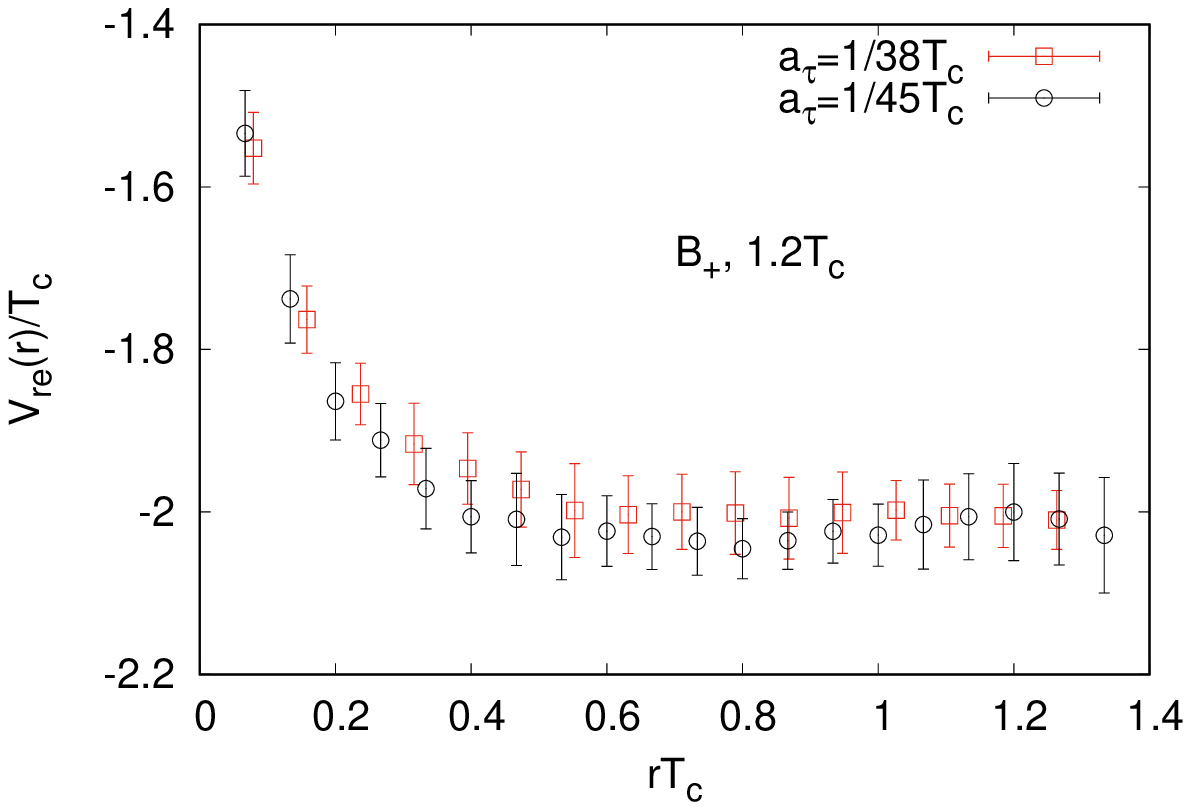}}
\centerline{\includegraphics[width=6.5cm]{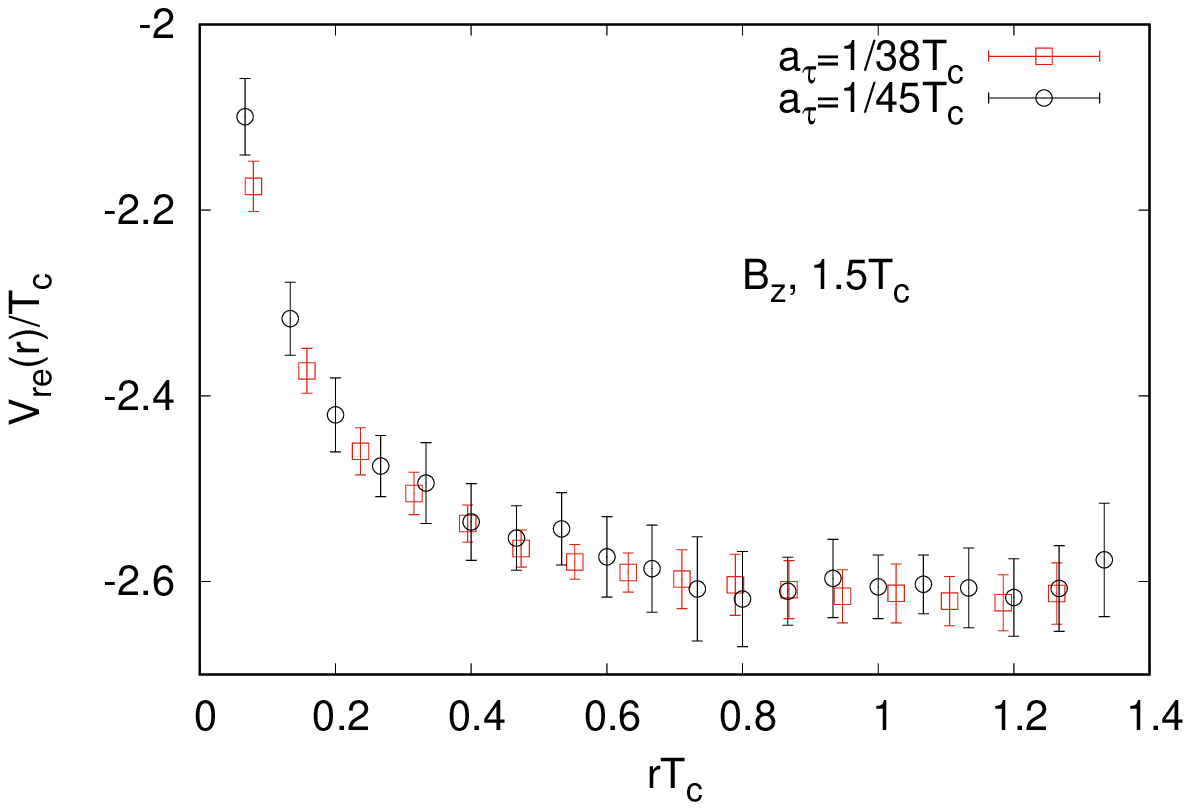}
  \includegraphics[width=6.5cm]{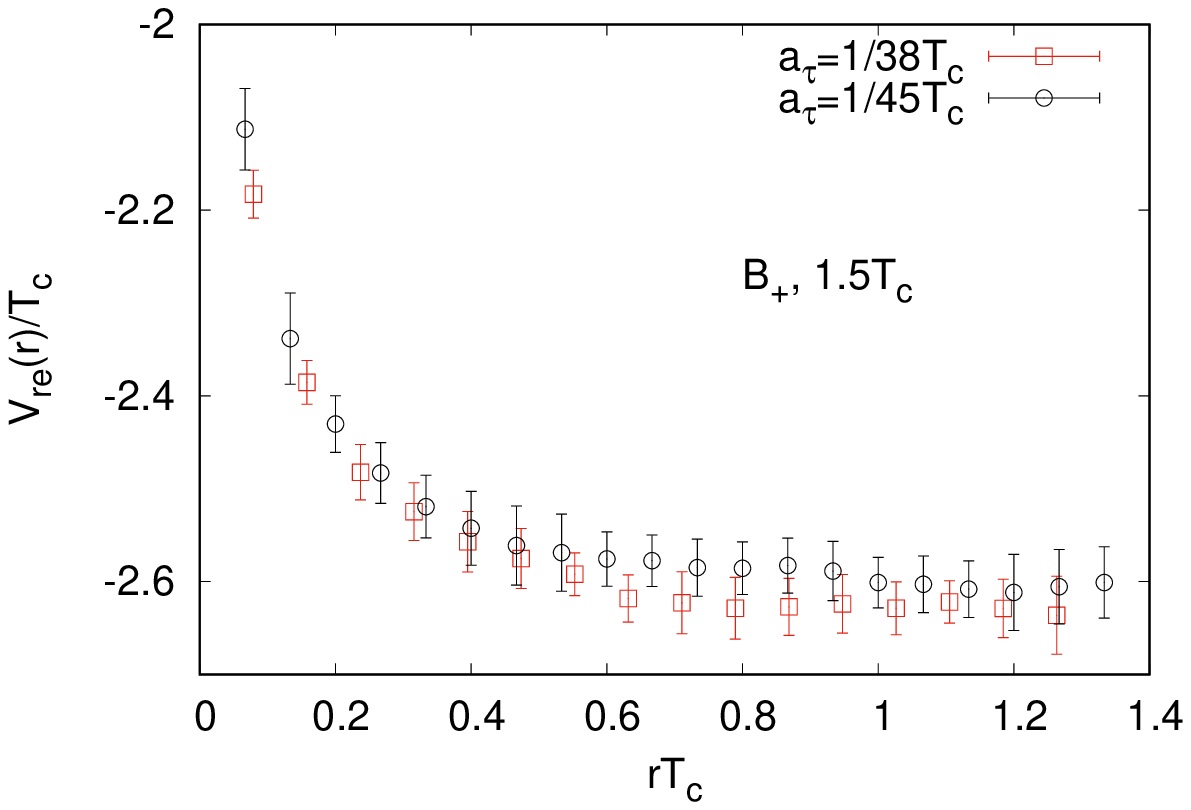}}
\caption{The results for $\vro$ extracted from lattices with different
  discretization levels. Shown are the results at 1.2 $\tc$ (top
  panels) and 1.5 $\tc$ (bottom panels). The left panels correspond to
  the L=0 channel while the right panels show the L=1 channel
  results.}
\eef{vr-cont}

\subsection{Systematics for $\vio$}
\label{sec.visystem}

For $\vio$ the smearing dependence of the extracted results is shown
in \fgn{vi-smear}. For the $\bz$ operator, the results stabilize
quickly: for set 3, already by 200 smearing steps the results seem to
have stabilized. For $\bxy$ it is similar, except at 1.2 $T_c$
at longer distances. Like for
$\vro$, the error bands quoted in \scn{vimag} include the spread
between smearing levels 300 and 400 for set 3, and that between
smearing levels 200 and 250 for set 2.

\bef
\centerline{\includegraphics[width=6.5cm]{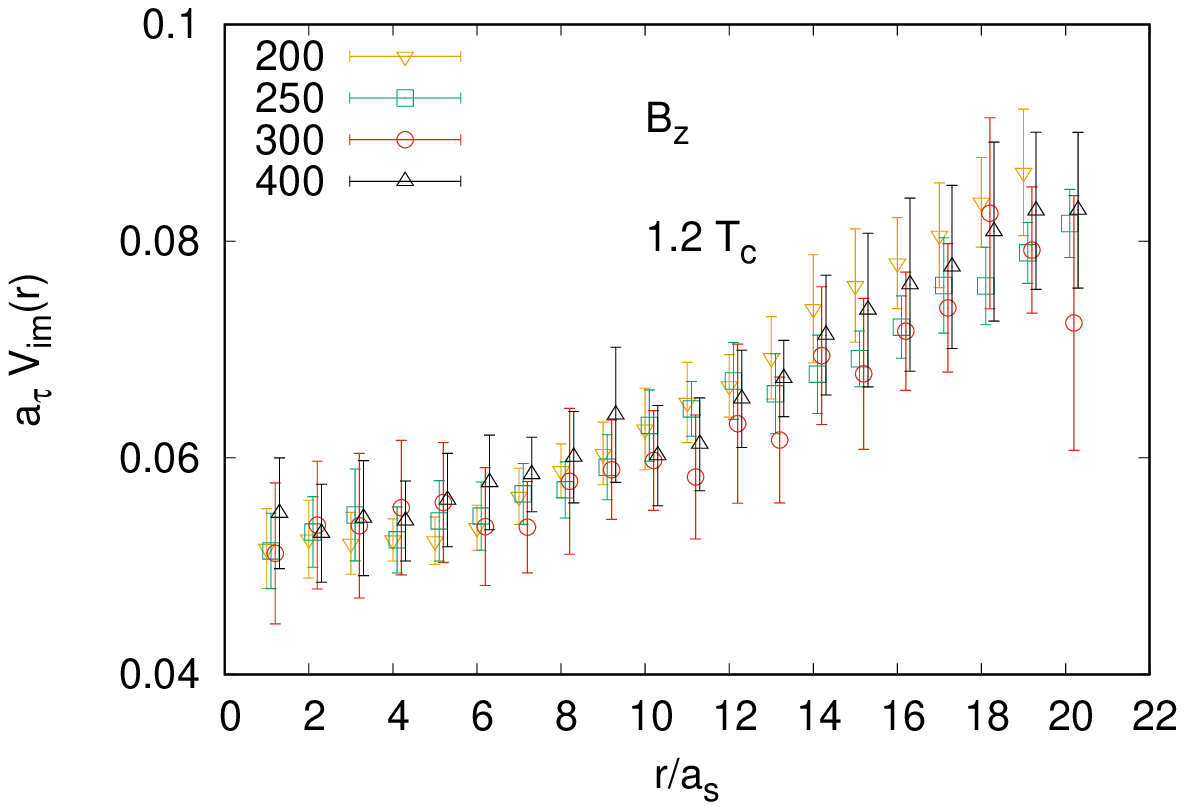}
  \includegraphics[width=6.5cm]{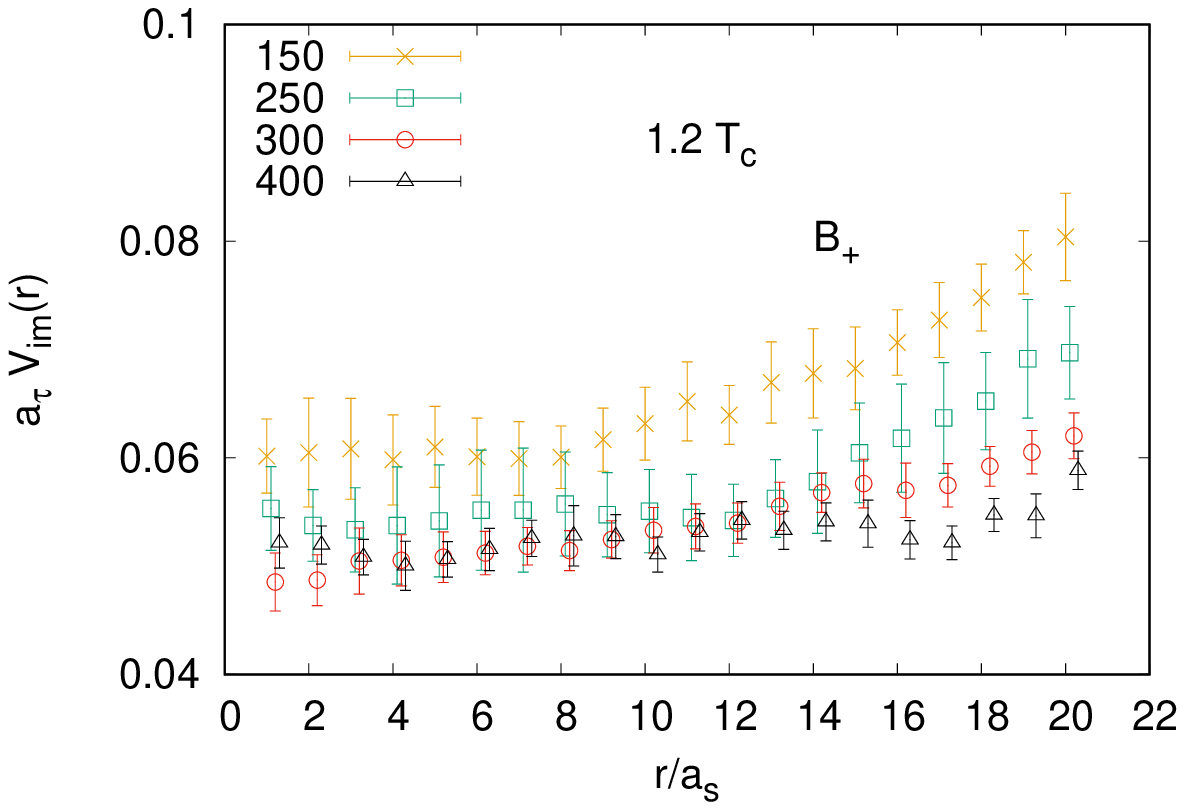}}
\centerline{\includegraphics[width=6.5cm]{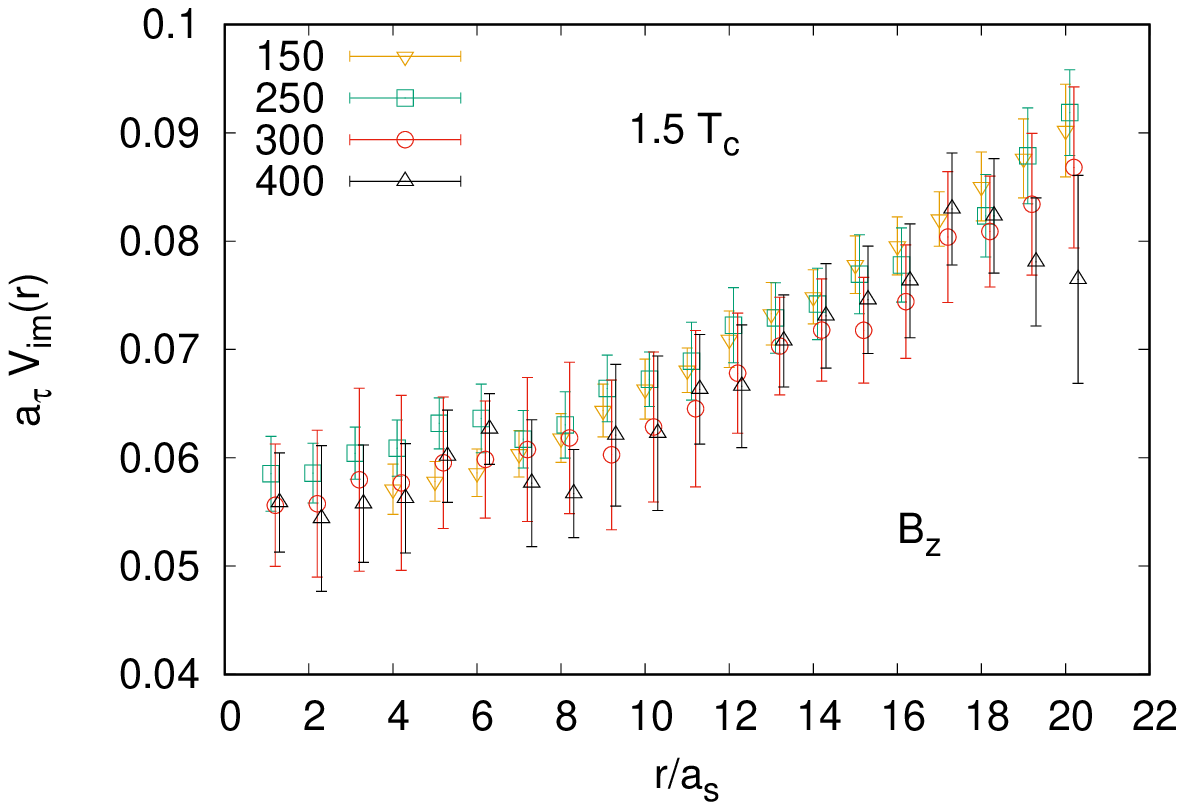}
  \includegraphics[width=6.5cm]{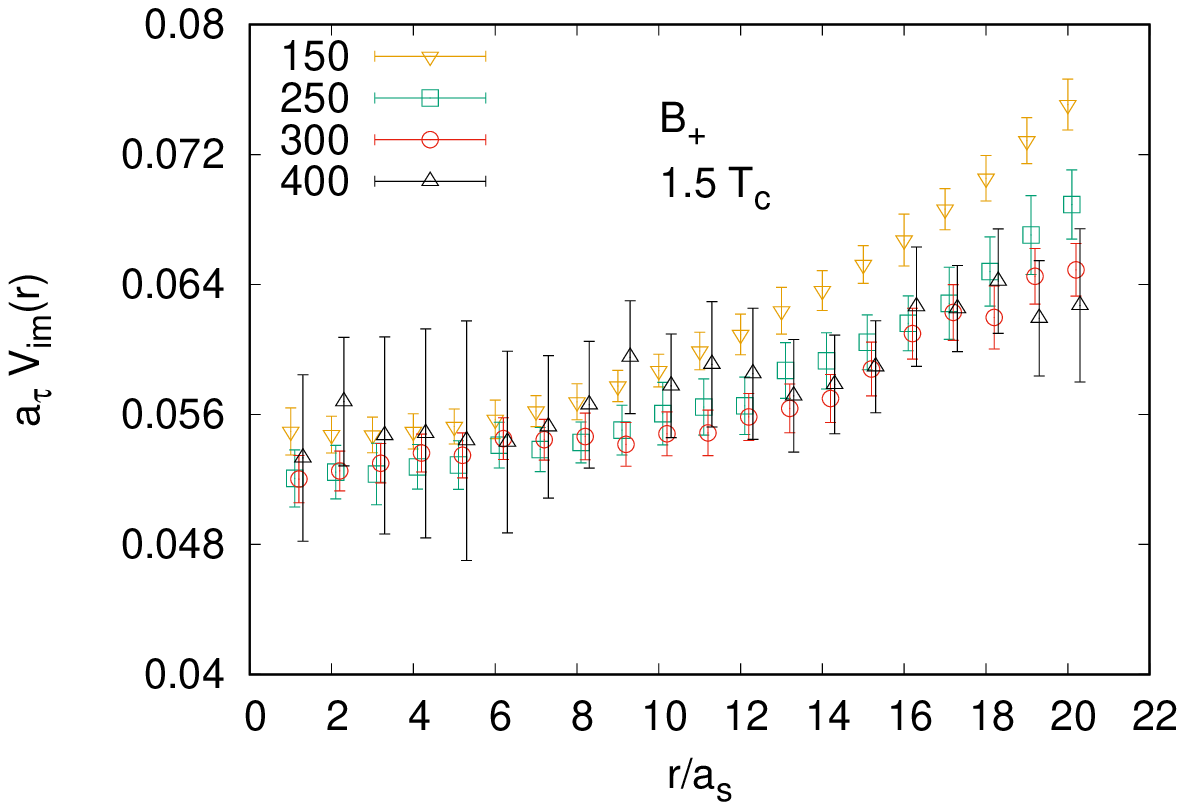}}
\caption{Smearing dependence of the imaginary part of the potential
  extracted from the hybrid operator inserted Wilson loops. The top
  row shows results at 1.2 $\tc$ while the bottom row shows results
  at 1.5 $\tc$. The panels to the left are for $\bz$ and those to
  the right are for $\bxy$ insertions. For viewing purpose some sets have
  been slightly shifted along x axis in the plot.}
\eef{vi-smear}

In comparison to the large errors associated with the extraction of
$\vio$, the cutoff effects do not seem to be significant. In
\fgn{vi-cont} we compare the results for $\vio$ from Set 1 and set 2.
Given the small cutoff dependence, we treat the results from our finest
lattices as indicative of continuum results. These are the results
shown in \scn{vimag}.

\bef
\centerline{\includegraphics[width=6.5cm]{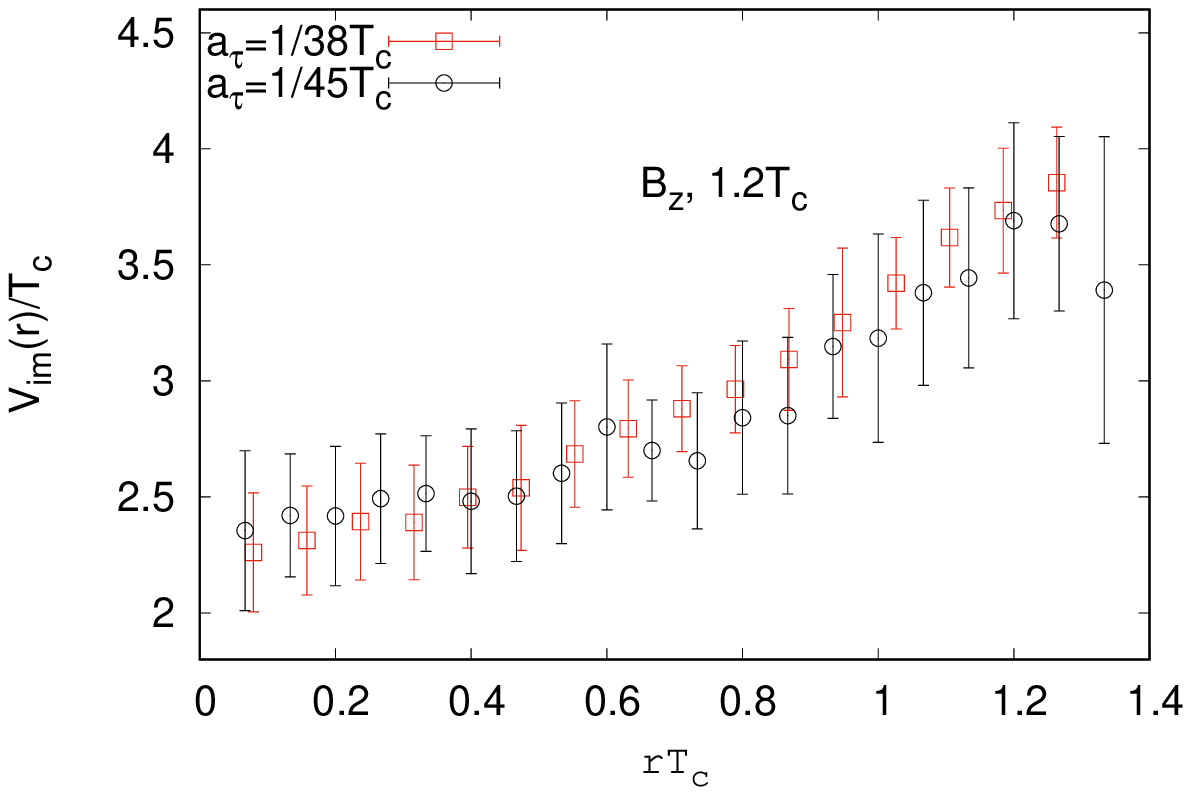}
  \includegraphics[width=6.5cm]{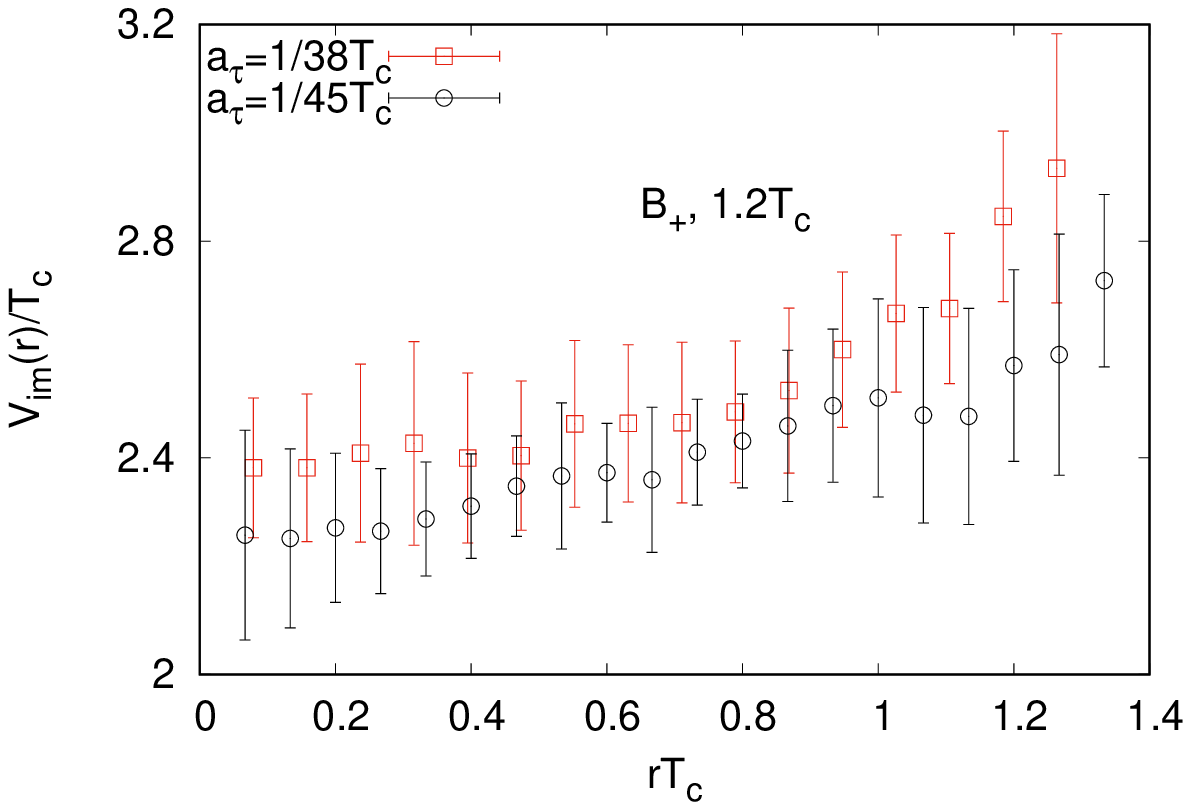}}
\centerline{\includegraphics[width=6.5cm]{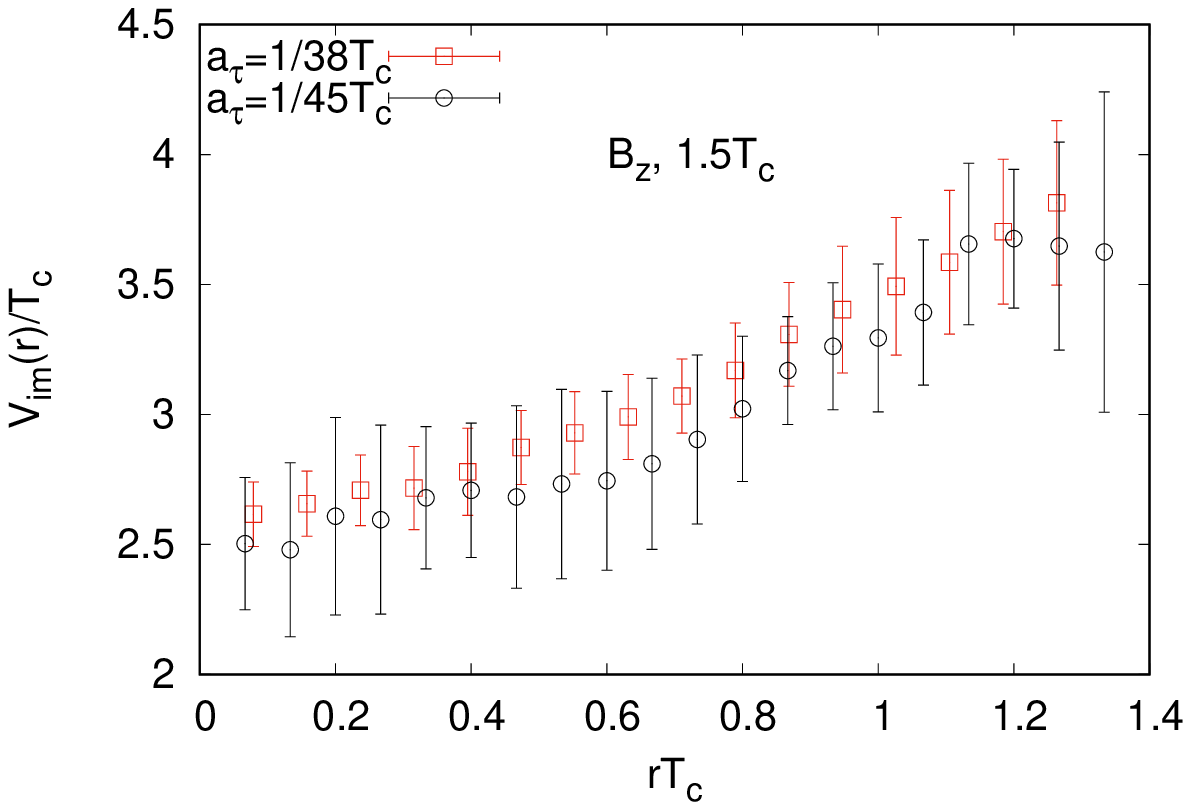}
  \includegraphics[width=6.5cm]{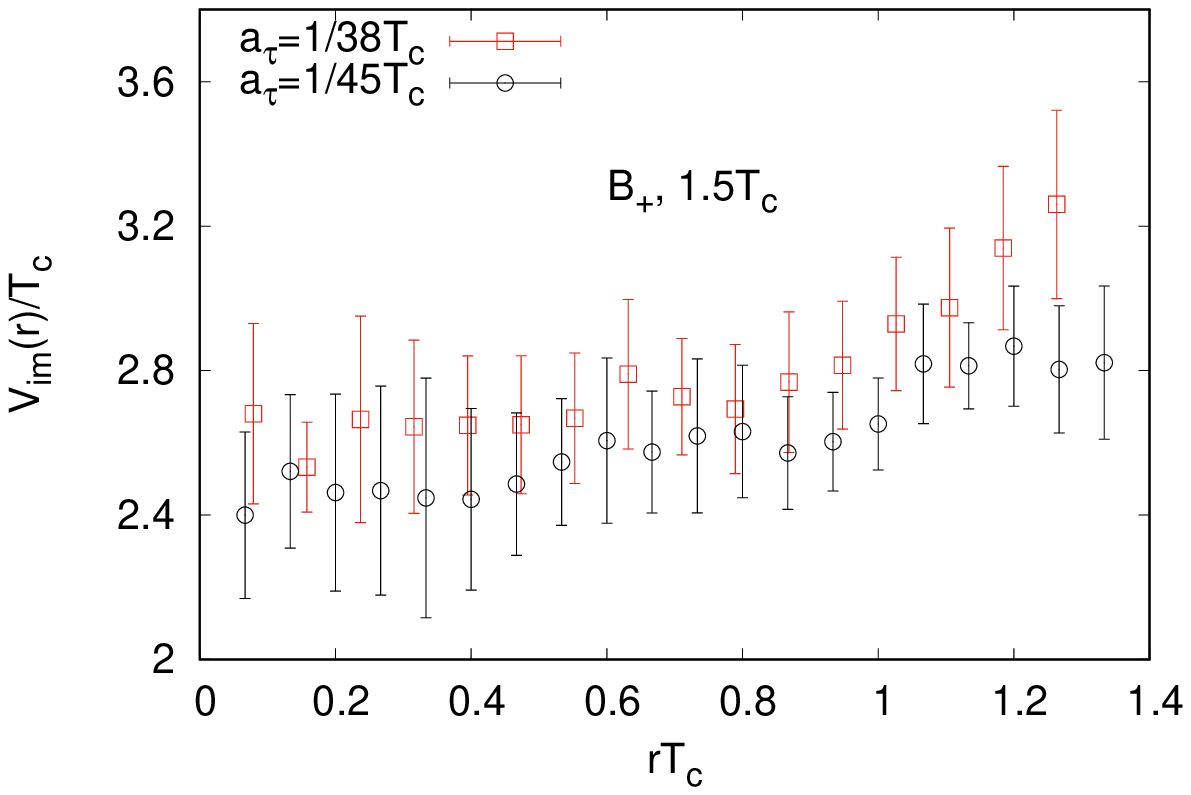}}
\caption{Comparison of results for $\vio$ at two different lattice
  spacings. The top row shows results at 1.2 $\tc$ while the bottom
  row shows results at 1.5 $\tc$. The panels to the left are for $\bz$
  and those to the right are for $\bxy$ insertions.}
\eef{vi-cont}

\input{bib.tex}
\end{document}

%% file: macros.tex
\newcommand\bef{\begin{figure}}
\newcommand\eef[1]{\label{fg:#1}\end{figure}}
\newcommand\beq{\begin{equation}}
\newcommand\eeq[1]{\label{#1}\end{equation}}
\newcommand\bea{\begin{eqnarray}}
\newcommand\eea[1]{\label{#1}\end{eqnarray}}
\newcommand\bet{\begin{table}}
\newcommand\eet[1]{\label{tbl:#1}\end{table}}

\newcommand\fgn[1]{Figure \ref{fg:#1}}
\newcommand\eqn[1]{Eq.\ (\ref{#1})}
\newcommand\scn[1]{Sec. \ref{sec.#1}}
\newcommand\apx[1]{Appendix \ref{sec.#1}}
\newcommand\tbn[1]{Table \ref{tbl:#1}}

\newcommand\jhep{{\sl J.\ H.\ E.\ P.\/}\ }
\newcommand\np{{\sl Nucl.\ Phys.\/}\ }

\newcommand\pr{{\sl Phys.\ Rev.\/}\ }
\newcommand\prlt{{\sl Phys.\ Rev.\ Lett.\/}\ }
\newcommand\plt{{\sl Phys.\ Lett.\/}\ }
\newcommand\prt{{\sl Phys.\ Rept.\/}\ }

\newcommand\ijmp{{\sl Int.\ J.\ Mod.\ Phys.}\ }

\newcommand{\mq}{M_{\scriptscriptstyle Q}}
\newcommand{\qqb}{Q \bar{Q}}

\newcommand{\qqo}{Q \bar{Q}\vert_o}
\newcommand{\cgt}{C_>(t, \vec{r})}
\newcommand{\cgto}{C_>^{\scriptscriptstyle J^a J^a}(t, \vec{r})}
\newcommand{\at}{a_\tau}
\newcommand{\as}{a_s}
\newcommand{\bs}{\beta_s}
\newcommand{\bt}{\beta_\tau}
\newcommand{\tc}{T_c}
\newcommand{\nt}{N_\tau}

\newcommand{\om}{\omega}

\newcommand{\sw}{\sigma(\omega; T)}
\newcommand{\vx}{\vec{x}}

\newcommand{\vxo}{\vec{x}_0}
\newcommand{\vrr}{\vec{r}}
\newcommand{\vt}{V_{\scriptscriptstyle T}(\vec{r})}
\newcommand{\nbw}{n_{\scriptscriptstyle B}(\omega)}

\newcommand{\viss}{V_{\rm im}}
\newcommand{\vrss}{V_{\rm re}}
\newcommand{\vre}{V^{\rm re}_{\scriptscriptstyle T}(\vec{r})}
\newcommand{\vim}{V^{\rm im}_{\scriptscriptstyle T}(\vec{r})}
\newcommand{\voc}{V^o(\vec{r}; T)}
\newcommand{\vrs}{V^s_{\rm re}(\vec{r}; T)}
\newcommand{\vis}{V^s_{\rm im}(\vec{r}; T)}
\newcommand{\vro}{V^o_{\rm re}(\vec{r}; T)}
\newcommand{\vio}{V^o_{\rm im}(\vec{r}; T)}
\newcommand{\md}{m_{\scriptscriptstyle D}}

\newcommand{\wrt}{W_{\scriptscriptstyle T}(\tau, \vec{r})}
\newcommand{\wrb}{W_{\scriptscriptstyle T}(\beta-\tau, \vec{r})}
\newcommand{\wpr}{W^p_{\scriptscriptstyle T}(\tau, \vec{r})}
\newcommand{\wap}{W^a_{\scriptscriptstyle T}(\tau, \vec{r})}
\newcommand{\ptau}{P(\tau)}
\newcommand{\wil}{\mathbb{U}}
\newcommand{\wm}{W_M}
\newcommand{\we}{W_E}
\newcommand{\wmg}{W_G}
\newcommand{\bz}{B_z}
\newcommand{\bxy}{B_+}
\newcommand{\rlow}{\rho_{{\rm low}}(r; \omega)}
\newcommand{\bose}{n_{\scriptscriptstyle B}}

\def\dtk(#1){\int \frac{d^3 {#1}}{8 \pi^3}}
\def\rhoe(#1){\rho_{\scriptscriptstyle E}({#1}_0, \vec{#1})}
\def\rhot(#1){\rho_{\scriptscriptstyle T}({#1}_0, \vec{#1})}
\def\pie(#1,#2){\Pi_{\scriptscriptstyle E}(\omega_{#1}, \vec{#2})}
\def\pit(#1,#2){\Pi_{\scriptscriptstyle T}(\omega_{#1}, \vec{#2})}
\def\intwh(#1){\int_{- \infty}^{\infty} \, \frac{d {#1}_0}{\pi}}
\def\ffk(#1){\mathcal{F}({#1}, \tau)}
\def\ggk(#1){\mathcal{G}({#1}, \tau)}
\def\ord(#1){\mathcal{O}({#1})}
\def\dvr(#1){\delta \, V^{\rm re}_{\scriptscriptstyle T}({#1})}